\documentclass[12pt]{article}

\usepackage{physics}
\usepackage{graphicx}
\usepackage{amssymb}
\usepackage{amsthm}
\usepackage{amsmath}
\usepackage{dsfont}
\usepackage{bm}
\usepackage{mathrsfs}
\usepackage{bbold}
\usepackage{color}
\usepackage{hyperref}
\usepackage{times}

\def\nn{\nonumber} 
\newcommand{\be}{\begin{equation}}
\newcommand{\ee}{\end{equation}}
\newcommand{\bea}{\setlength\arraycolsep{2pt} \begin{eqnarray}}
\newcommand{\eea}{\end{eqnarray}}

\title{\large {\bf Extracting the expression for the field equations of
a diffeomorphism invariant theory of gravity from surface term}}
\date{}

\author{Jun-Jin Peng\footnote{corresponding author: pengjjph@163.com}
   \\
\small \sl School of Physics and Electronic Science,
\small \sl Guizhou Normal University,\\
\small Guiyang, Guizhou 550025, People's Republic of China \\
}

\begin{document}
\maketitle
\vspace{-10pt}

\begin{center}
\textbf{Abstract}
\end{center}
\vspace{5pt}

As a contribution towards the understanding for the field equations
of diffeomorphism invariant theories of pure gravity, we demonstrate
in great detail that the expression for the field equations of such
theories can be derived within the perspective of the surface term
coming from the variation of the Lagrangian. Specifically, starting
with the surface term, we extract a symmetric rank-two tensor together
with an anti-symmetric one out of this term with the variation
operator replaced with the Lie derivative along an arbitrary vector
field. By utilizing an equality stemming from the Lie derivative of the
Lagrangian density along an arbitrary vector field, it is proved that the
resulting symmetric rank-two tensor is identified with the functional
derivative of the Lagrangian density with respect to the metric. Such
a result further brings forth the expression for the field equations
constructed from the symmetric rank-two tensor, which naturally rules
out the derivative of the Lagrangian density with respect to the metric
and coincides with the one for the Euler-Lagrange equations of motion.
Furthermore, it is illustrated that the construction of the expression
for the field equations from the symmetric rank-two tensor must be
feasible as long as the variation operator in the variation equation of
the Lagrangian is allowed to be substituted by the Lie derivative along
an arbitrary vector field. On the other hand, as a byproduct, the
anti-symmetric rank-two tensor turns out to be the Noether charge
two-form. Subsequently, it is illustrated in detail
that there exist two equivalent ways to identify the field equations
arising from the surface term with the Euler-Lagrange equations.
In addition, utilizing the surface term, together with two equalities
derived from the variation equations of the Lagrangian with the
variation operator substituted by the Lie derivative, we obtain some
equivalence relationships associated to the conserved current and the
expression for the field equations. Our results offer a straightforward
support on the proposal in our previous work that the surface term
gives a unified description for field equations and Noether charges in
the context of theories of gravity admitting diffeomorphism invariance
symmetry. What is more, we observe that the expression for the field
equations can appear as a consequence for the vanishing of its
divergence or the divergence for the off-shell Noether current.


\voffset=-.90pt
\vspace{10pt}
\newpage
\section{Introduction}\label{one}

Given a Lagrangian based theory of gravity, the conventional approach
to derive the gravitational field equations associated to this theory
is to vary the Lagrangian with respect to the metric under the guidance
of the variational principle, which is a powerful tool in field theories
\cite{IyerWald,BarnichB,DHint09,PetPit,FGie21}. From a general
perspective, the resulting equations of motion, referred to as the
Euler-Lagrange equations in the literature as well, are expressed
through the functional derivative of the Lagrangian with respect to
the metric. This expression ordinarily gets involved in the derivative
of the Lagrangian density with respect to the metric. Thus, the
elimination of such a term renders it of possibility to simplify the
equations of motion.

Within the paper \cite{Pady}, by utilizing an identity establishing
the relation between the derivatives of the Lagrangian density with
respect to the metric and the Riemann curvature tensor respectively,
which appears as a consequence of the comparison between the results
for the Lie derivative of the Lagrangian density evaluated in two
different ways, Padmanabhan presented a compact form for the field
equations of gravity without the derivative of the Lagrangian density
with respect to the metric in the context of generally covariant metric
theories. However, since the Lagrangians under consideration merely
depend upon the metric and the Riemann curvature tensor, the
application range of the resulting field equations is restricted to
gravity theories equipped with the Lagrangian that is free of the
covariant derivative(s) of the Riemann tensor. Subsequently, as a
generalization to the higher derivative theories of gravity
constructed out of the first-order covariant derivative of the
Riemann tensor in addition to the metric and the Riemann tensor,
Padmanabhan's method was extended to derive the equations
of motion for such theories in the paper \cite{ERR22}. Another
generalization to the gravity theories in the presence of a vector
field together with its first-order covariant derivative was given by
the paper \cite{WallY24} quite recently. Till now, owing to the
conciseness of the field equations presented by \cite{Pady}, they
have gained a broad range of applications. Some of them can be found
in the works \cite{LYu25,GGLMZ24,PWG23,CRT22,Mur21,AESRL21,Maed20,
BCMV16,DLM16,GJS12,CPad15}.

What is more, for a generally covariant Lagrangian-based metric
theory of gravity with or without the covariant derivatives of
the Riemann tensor, the variation of the Lagrangian with regard to
the metric brings forth a surface term that naturally eliminates
the derivative of the Lagrangian density with respect to the metric.
As is what has been illustrated in our earlier work \cite{JJPEQM},
such a surface term, which is also referred to as the symplectic
potential current density \cite{LeeWald},
can be employed to generate the expression for the field equations
of the gravitational field. Not unnaturally, throughout the whole
process for doing so, one is able to keep away from any calculation
involving the derivative of the Lagrangian density with respect to
the metric. By contrast, in the framework of Padmanabhan's method,
it is unavoidable to deal with the identity associated to this
quantity as it plays a prominent role in simplifying the equations of
motion fully arising from the variation of the Lagrangian. Apparently,
the field equations from the perspective of the surface term
naturally inherit the feature for the surface term, that is, the
derivative of the Lagrangian density with respect to the metric
is completely excluded from them.

According to the approach put forward in Ref. \cite{JJPEQM},
the surface term stemming from the variation of the Lagrangian plays
a central role. Under the substitution for the variation operator by
the Lie derivative along an arbitrary vector field, the resulting surface
term can be decomposed into the linear combination of two components.
One of them is the contraction between the vector field and a symmetric
rank-two tensor, while the other consists of the covariant divergence of
another anti-symmetric rank-two tensor. It can be verified that the symmetric
rank-two tensor corresponds to the expression for the field equations
by making use of an identity linking its divergence with that of an
off-shell Noether current. Actually, the approach to produce the field
equations based on the surface term provides a natural framework for
unifying both the field equations and the conserved current. Apart from
the extraction of the field equations, the Noether charge two-form,
whose Hodge dual is referred to as Noether charge $(D-2)$-form in
$D$-dimensional spacetimes within the papers \cite{IyerWald,NoeChWald},
is able to be produced simultaneously. In practice, it is just the
anti-symmetric rank-two tensor encompassed by the
surface term. As an application, this surface term based approach has
been extended to the higher derivative theories of gravity endowed
with Lagrangians built from the metric and the Riemann tensor, together
with $i$th ($i=1,2,\cdot\cdot\cdot$) powers of the
Beltrami-d'Alembertian operator acting on the latter \cite{PL2402}.
In addition, utilizing the field equations presented by \cite{JJPEQM},
the authors investigated the stellar axisymmetry theorem in the context
of a broad class of metric theories of gravity admitting diffeomorphism
invariance within the paper \cite{DKS2512}.

The present paper is an extension and development to our earlier work
\cite{JJPEQM}. It mainly focuses on some fundamental problems on
how to extract the expression for the field equations out of the
surface term arising from the variation of the Lagrangian, such as
whether the surface term must accommodate the rank-two symmetric tensor
that is able to participate in the construction of the field equations,
or what constitute the foundation for the derivation of the field
equations from the surface term, or what characterize the expression
for the equations of motion extracted out of the surface term. Regarding
these issues, this paper encompasses four main objectives. First,
building upon the work \cite{JJPEQM}, we attempt to seek another route
to confirming that the expression of the field equations for gravity
theories admitting diffeomorphism invariance symmetry can be extracted
out of the surface term. On the basis of the two routes, our intention
is to establish the sufficient and necessary conditions for that the
surface term is able to produce such an expression. Thus it will be
verified that deriving field equations from the surface term perspective
indeed establishes on a solid foundation. Due to this, we expect that
our approach will enhance the understanding of the implications for
utilizing the symmetries of the Lagrangian to yield the equations of
motion. Second, due to that the surface term is able to provide a
unified description towards the conserved Noether current and the
expression for the field equations, we anticipate giving some insights
into the relationships between both of them. Furthermore, inspired with
the results, we plan to verify that the expression for the equations
of motion can stem from the disappearance of its covariant divergence.
Third, in terms of the surface term perspective towards the derivation
for the expression of the field equations, we observe that three equations
(see Eqs. (\ref{ThetaLie}), (\ref{LieLagdens}) and (\ref{DivJE}) below
for them), which result from performing the replacement of the variation
operator with the Lie derivative towards the surface term and the
variation equations of the Lagrangian, occupy a decisive position in
producing the desired expression for the equations of motion. Thus,
we plan to reveal the deep connections among them. Fourth, within the
context of the diffeomorphism invariant theories of gravity comprising
arbitrary order covariant derivatives of the Riemann tensor, under the
guidance of the expression for the field equations within the surface
term perspective, we attempt to fill the gap in the derivation for the
field equations of these theories by strictly following Padmanabhan's
method. At the same time, we will carry out a detailed comparison
between the two approaches in order to offer an improved understanding
of Padmanabhan's method. In addition, we intend to obtain the explicit
Noether charge two-forms for these theories.

The present paper is structured as follows. Within Section \ref{two},
the equations of motion and the surface term for diffeomorphism
invariant theories of pure gravity involving the covariant
derivatives of the Riemann tensor are figured out via the variation of
the Lagrangians for these theories. Within Section \ref{three},
under the replacement for the variation operator by the Lie derivative
along an arbitrary smooth vector field, the surface term is put into
the form consisting of a symmetric rank-two tensor and an anti-symmetric
one. By using an equality associated to the partial derivative of the
Lagrangian density with respect to the metric, which is derived out of
the Lie derivative of the Lagrangian density, then it is demonstrated that
the symmetric rank-two tensor gives rise to the field equations and the
anti-symmetric rank-two tensor represents the Noether charge two-form
corresponding to the Noether current. Within Section \ref{four}, we make
a detailed comparison between two different ways for identifying the
expression for the Euler-Lagrange equations of motion with the one
arising from the surface term. By the way, we obtain some equivalence
relations relative to the conserved current and the expression for the
field equations. Within Section \ref{five}, we reveal several relationships
among three significant equations arising from the replacement of the
variation operator with the Lie derivative along an arbitrary vector
field in the surface term and the variation equations of the Lagrangian.
Along the way, we explore the sufficient and necessary condition for
the derivation of the field equations from the surface term.
Within Section \ref{six}, through straightforward calculations,
it is proved that the expression for the field equations originating
from the surface term is divergence-free. Within Section \ref{seven},
in terms of all the aforementioned analysis, the procedure for the
derivation of the field equations from the surface term perspective
is put forward. According to the procedure, we derive the explicit
expressions for the surface term and the field equations associated
to the generic theories of gravity endowed with the Lagrangian
relying on the metric, the Riemann tensor and the covariant derivative
operator. This section encompasses the main idea of the present work.
The last section contains our conclusions and discussions.

\section{Euler-Lagrange equations and surface terms resulting from
the variation of the Lagrangian}\label{two}

In this section, we are going to derive field equations
and surface terms by means of varying the Lagrangian with
regard to the metric tensor.

As a beginning, we take into account a generic Lagrangian in arbitrary
dimensions that is invariant under a change of coordinates on the
spacetime manifold. Specifically, this diffeomorphism invariant Lagrangian
is supposed to be built out of the metric together with $i$th-order
$(i=0,1,\cdot\cdot\cdot,m)$ covariant derivatives
of the Riemann curvature tensor. It has the form \cite{IyerWald}
\be
\mathcal{L}=\sqrt{-g}L\left(g^{\mu\nu},
R_{\mu\nu\rho\sigma},\nabla_{\lambda_1}R_{\mu\nu\rho\sigma},
\cdot\cdot\cdot,\nabla_{\lambda_1}\cdot\cdot\cdot
\nabla_{\lambda_m}R_{\mu\nu\rho\sigma}\right)
\, . \label{LagCovR}
\ee
Computing straightforwardly the variation of the Lagrangian density
$L$ with respect to all of its constituent variables, that is,
the inverse metric $g^{\mu\nu}$ and the covariant derivatives of the
Riemann curvature tensor $\nabla_{\lambda_1}\cdot\cdot\cdot
\nabla_{\lambda_i}R_{\mu\nu\rho\sigma}$s, we obtain the result
\be
\delta{L}=\frac{\partial{L}}{\partial{g}^{\mu\nu}}\delta{g}^{\mu\nu}
+\sum^{m}_{i=0}
Z_{(i)}^{\lambda_1\cdot\cdot\cdot\lambda_i\mu\nu\rho\sigma}
\delta\nabla_{\lambda_1}
\cdot\cdot\cdot\nabla_{\lambda_i}R_{\mu\nu\rho\sigma}
\, , \label{VariLagdens}
\ee
where the rank-$(i+4)$ tensor
$Z_{(i)}^{\lambda_1\cdot\cdot\cdot\lambda_i\mu\nu\rho\sigma}$,
standing for the derivative of the Lagrangian density with respect
to the variable $\nabla_{\lambda_1}\cdot\cdot\cdot
\nabla_{\lambda_i}R_{\mu\nu\rho\sigma}$, is presented by
\be
Z_{(i)}^{\lambda_1\cdot\cdot\cdot\lambda_i\mu\nu\rho\sigma}
=Z_{(i)}^{\lambda_1\cdot\cdot\cdot\lambda_i\alpha\beta\gamma\lambda}
\Delta^{\mu\nu\rho\sigma}_{\alpha\beta\gamma\lambda}
=\frac{\partial{L}}{\partial
\nabla_{\lambda_1}\cdot\cdot\cdot
\nabla_{\lambda_i}R_{\mu\nu\rho\sigma}}
\, . \label{Zidef}
\ee
The $\Delta^{\mu\nu\rho\sigma}_{\alpha\beta\gamma\lambda}$ tensor
in Eq. (\ref{Zidef}), employed to characterize the same index
symmetries as those for the Riemann tensor,
is defined in terms of the Kronecker delta through
\be
\Delta^{\mu\nu\rho\sigma}_{\alpha\beta\gamma\lambda}
=\frac{1}{2}\left(\delta^{[\mu}_{\alpha}\delta^{\nu]}_{\beta}
\delta^{[\rho}_{\gamma}\delta^{\sigma]}_{\lambda}
+\delta^{[\rho}_{\alpha}\delta^{\sigma]}_{\beta}
\delta^{[\mu}_{\gamma}\delta^{\nu]}_{\lambda}\right)
\, . \label{Delttdef}
\ee
Within Eqs. (\ref{VariLagdens}) and (\ref{Zidef}), it is to be
understood that no covariant derivatives act on the Riemann tensor
when $i=0$ and
$Z_{(i)}^{\lambda_1\cdot\cdot\cdot\lambda_i\mu\nu\rho\sigma}$ with
$i=0$ represents $Z_{(0)}^{\mu\nu\rho\sigma}=
\frac{\partial{L}}{\partial{R}_{\mu\nu\rho\sigma}}$.
In terms of Eq. (\ref{VariLagdens}), the variation of the Lagrangian
(\ref{LagCovR}) is written down as
\be
\delta\mathcal{L}
=\sqrt{-g}\left(\delta{L}
-\frac{1}{2}{L}{g}_{\mu\nu}\delta{g}^{\mu\nu}\right)
\, . \label{VarLag}
\ee
For convenience, throughout this work, we introduce a third-rank tensor
${U}^{\mu\alpha\beta}_{(i)}$ defined
in terms of $Z_{(i)}^{\lambda_1\cdot\cdot\cdot\lambda_i\mu\nu\rho\sigma}$s
through
\bea
{U}^{\mu\alpha\beta}_{(i)}&=&
4\sum^i_{k=1}(-1)^k\Big(\nabla_{\lambda_{k-1}}
\cdot\cdot\cdot\nabla_{\lambda_1}
Z_{(i)}^{\lambda_1\cdot\cdot\cdot
\lambda_{k-1}\alpha\lambda_{k+1}\cdot\cdot\cdot
\lambda_{i}\beta\gamma\rho\sigma}\Big)
\nabla_{\lambda_{k+1}}\cdot\cdot\cdot\nabla_{\lambda_i}
R^\mu_{~\gamma\rho\sigma}
\nn \\
&&+\sum^{i-1}_{k=1}\sum^i_{\ell=k+1}(-1)^k
\Big(\nabla_{\lambda_{k-1}}
\cdot\cdot\cdot\nabla_{\lambda_1}
Z_{(i)}^{\lambda_1\cdot\cdot\cdot\lambda_{k-1}
\alpha\lambda_{k+1}\cdot\cdot\cdot
\lambda_{\ell-1}\beta\lambda_{\ell+1}
\cdot\cdot\cdot\lambda_{i}
\gamma\lambda\rho\sigma}\Big)\times \nn \\
&&\times \nabla_{\lambda_{k+1}}\cdot\cdot\cdot
\nabla_{\lambda_{\ell-1}}\nabla^\mu
\nabla_{\lambda_{\ell+1}}\cdot\cdot\cdot
\nabla_{\lambda_{i}}R_{\gamma\lambda\rho\sigma}
\, . \label{Uidef}
\eea
For instance, in the above equation, ${U}^{\mu\alpha\beta}_{(0)}=0$
and ${U}^{\mu\alpha\beta}_{(1)}=-4Z_{(1)}^{\alpha\beta\gamma\rho\sigma}
R^\mu_{~\gamma\rho\sigma}$.
By the aid of the third-rank tensor ${U}^{\mu\alpha\beta}_{(i)}$,
the scalar $Z_{(i)}^{\lambda_1\cdot\cdot\cdot\lambda_i\mu\nu\rho\sigma}
\delta \nabla_{\lambda_1}
\cdot\cdot\cdot\nabla_{\lambda_i}R_{\mu\nu\rho\sigma}$ can be
put into the form
\bea
Z_{(i)}^{\lambda_1\cdot\cdot\cdot\lambda_i\mu\nu\rho\sigma}
\delta \nabla_{\lambda_1}
\cdot\cdot\cdot\nabla_{\lambda_i}{R}_{\mu\nu\rho\sigma}&=&
(-1)^i\big(\nabla_{\lambda_{i}}\cdot\cdot\cdot\nabla_{\lambda_1}
Z_{(i)}^{\lambda_1\cdot\cdot\cdot\lambda_i\mu\nu\rho\sigma}\big)
\delta{R}_{\mu\nu\rho\sigma} \nn \\
&&-{B}^{\mu\nu}_{(i)}\delta{g}_{\mu\nu}
+\nabla_\mu\bar{\Theta}^\mu_{(i)}
\, . \label{ZidelnabR}
\eea
Within Eq. (\ref{ZidelnabR}), the second-rank symmetric tensor
${B}^{\mu\nu}_{(i)}$ is defined in terms of the divergences
of the tensor ${U}^{\mu\alpha\beta}_{(i)}$ as
\be
{B}^{\mu\nu}_{(i)}=\frac{1}{2}\nabla_\lambda
\left({U}^{(\mu\nu)\lambda}_{(i)}
-{U}^{\lambda(\mu\nu)}_{(i)}
+{U}^{(\mu|\lambda|\nu)}_{(i)}\right)
\, , \label{Bimndef}
\ee
while the vector $\bar{\Theta}^\mu_{(i)}$, appearing as a result of
peeling off the covariant derivative operator in
$\delta \nabla_{\lambda_1}\cdot\cdot\cdot\nabla_{\lambda_i}
{R}_{\mu\nu\rho\sigma}$, has the form
\bea
\bar{\Theta}^\mu_{(i)}&=&
\sum^i_{k=1}(-1)^{k-1}\Big(\nabla_{\lambda_{k-1}}\cdot\cdot\cdot
\nabla_{\lambda_{1}}
{Z}^{\lambda_1\cdot\cdot\cdot\lambda_{k-1}\mu
\lambda_{k+1}\cdot\cdot\cdot\lambda_{i}\alpha\beta\rho\sigma}_{(i)}
\Big) \delta \nabla_{\lambda_{k+1}}
\cdot\cdot\cdot\nabla_{\lambda_{i}}{R}_{\alpha\beta\rho\sigma}\nn \\
&&+\frac{1}{2}\left({U}^{(\alpha|\mu|\beta)}_{(i)}
+{U}^{(\alpha\beta)\mu}_{(i)}
-{U}^{\mu(\alpha\beta)}_{(i)}\right)\delta {g}_{\alpha\beta}
\, . \label{Thetbar}
\eea
Particularly, $\bar{\Theta}^\mu_{(0)}=0$. Obviously, the dependence of
$\bar{\Theta}^\mu_{(i)}$ on
$\delta \nabla_{\lambda_1}\cdot\cdot\cdot\nabla_{\lambda_i}
{R}_{\mu\nu\rho\sigma}$ and $\delta {g}_{\alpha\beta}$ is linear.
As a consequence of Eq. (\ref{ZidelnabR}), equation (\ref{VarLag})
is recast into
\bea
\delta\mathcal{L}
&=&\sqrt{-g}\left[\left(\frac{\partial{L}}{\partial {g}^{\mu\nu}}
-\frac{1}{2}{L} {g}_{\mu\nu}+{B}_{\mu\nu}\right)
\delta g^{\mu\nu}
+{P}^{\mu\nu\rho\sigma}\delta{R}_{\mu\nu\rho\sigma} \right]\nn \\
&&+\sqrt{-g}\nabla_\mu\left(\sum^{m}_{i=0}\bar{\Theta}^\mu_{(i)}\right)
\, . \label{VarLag2}
\eea
In Eq. (\ref{VarLag2}), the second-rank symmetric tensor
${B}^{\mu\nu}$ is defined as
\be
{B}^{\mu\nu}=\sum^m_{i=0}{B}^{\mu\nu}_{(i)}
\, , \label{Bmndef}
\ee
and the rank-four tensor ${P}^{\mu\nu\rho\sigma}$
is in the form
\be
{P}^{\mu\nu\rho\sigma}=\sum^{m}_{i=0}
(-1)^{i}
\nabla_{\lambda_{i}}\cdot\cdot\cdot\nabla_{\lambda_1}
Z_{(i)}^{\lambda_1\cdot\cdot\cdot\lambda_i\mu\nu\rho\sigma}
\, . \label{Pdef}
\ee

Furthermore, after utilizing \cite{JJPEQM}
\be
{P}^{\mu\nu\rho\sigma}\delta{R}_{\mu\nu\rho\sigma}
=-\big({P}_{\mu}^{~\lambda\rho\sigma}R_{\nu\lambda\rho\sigma}
+2\nabla^\rho\nabla^\sigma{P}_{\rho\mu\nu\sigma}\big)
\delta{g}^{\mu\nu}+\nabla_\mu\Theta^\mu_{P}
\, , \label{PdelRiem}
\ee
with the vector $\Theta^\mu_{P}$ given by
\be
\Theta^\mu_{P}
=2{P}^{\mu\nu\rho\sigma}\nabla_\sigma\delta{g}_{\rho\nu}
-2(\delta{g}_{\nu\rho})\nabla_\sigma{P}^{\mu\nu\rho\sigma}
\, ,\label{ThetPdef}
\ee
we find that equation (\ref{VarLag2}) is ultimately written as
the conventional form for the variation of the Lagrangian
\cite{IyerWald,LeeWald,NoeChWald}, namely,
\be
\delta\mathcal{L}=\sqrt{-g}\left({E}_{\mu\nu}\delta{g}^{\mu\nu}
+\nabla_\mu\Theta^\mu\right)
\, . \label{VarLag3}
\ee
Within the above equation, the expression ${E}_{\mu\nu}$ for equations
of motion is read off as
\bea
{E}_{\mu\nu}&=&\frac{1}{\sqrt{-g}}
\frac{\delta\mathcal{L}}{\delta{g}^{\mu\nu}}
=\frac{\delta{L}}{\delta{g}^{\mu\nu}}
-\frac{1}{2}{L} {g}_{\mu\nu}\nn \\
&=&\frac{\partial{L}}{\partial {g}^{\mu\nu}}
-\frac{1}{2}{L} {g}_{\mu\nu}+{B}_{\mu\nu}
-{P}_{\mu}^{~\lambda\rho\sigma}R_{\nu\lambda\rho\sigma}
-2\nabla^\rho\nabla^\sigma{P}_{\rho\mu\nu\sigma}
\, . \label{EulerLagEq}
\eea
In terms of Eq. (\ref{EulerLagEq}), the field equations, that is,
the Euler-Lagrange equations of motion, are written down as
\be
{E}_{\mu\nu}=0
\, . \label{EoMeqzero}
\ee
The surface term $\Theta^\mu$ in Eq. (\ref{VarLag3}) incorporates
$\Theta^\mu_{P}$ and $\bar{\Theta}^\mu_{(i)}$s in the way
\be
\Theta^\mu=\Theta^\mu_{P}
+\sum^{m}_{i=0}\bar{\Theta}^\mu_{(i)}
\, , \label{SurfT}
\ee
which fulfills Eq. (\ref{SumZideltR}). Apparently, here the surface
term $\Theta^\mu$ merely arises from the
$Z_{(i)}^{\lambda_1\cdot\cdot\cdot\lambda_i\mu\nu\rho\sigma}
\delta\nabla_{\lambda_1}
\cdot\cdot\cdot\nabla_{\lambda_i}R_{\mu\nu\rho\sigma}$ terms within
the variation equation (\ref{VariLagdens}) for the Lagrangian density
and it naturally rules out the symmetric tensor
$\frac{\partial{L}}{\partial{g}^{\mu\nu}}$
or $\frac{\partial{L}}{\partial {g}_{\mu\nu}}$. This accordingly
causes the vanishing of such a tensor in the field equations resulting
from the surface term as is what will be shown in the next section.
Like usual, the expression (\ref{EulerLagEq}) for the equations of motion
is obtained by following the variational principle to directly vary
the Lagrangian with respect to the metric. It adequately comes
from the terms proportional to the variation of the metric and
is irrelevant to the surface terms. This implies that the surface term
can be ignored if the purpose is only to derive the Euler-Lagrange
equations for gravitational field. However, it will be illustrated
below that the surface term is of great importance throughout our
analysis. It encodes all the information for the expressions of
the field equations and the conserved Noether current.
In addition, owing to the fact that both the
second-rank tensors $\frac{\partial{L}}{\partial {g}^{\mu\nu}}$ and
${B}_{\mu\nu}$ are symmetric, together with the equality
\cite{Pady,PWG23}
\be
{P}_{[\mu}^{~~\lambda\rho\sigma}R_{\nu]\lambda\rho\sigma}
+2\nabla^\rho\nabla^\sigma{P}_{\rho[\mu\nu]\sigma}=0
\, , \label{PRantSym}
\ee
it is always guaranteed that ${E}_{\mu\nu}={E}_{\nu\mu}$.

\section{Identifying the expression for the field equations arising
from surface term with the Euler-Lagrange expression}\label{three}

In this section, we deal with the surface term under the transformation
of the variation operator to the Lie derivative along an arbitrary
smooth vector field. From the surface term via such a transformation,
a rank-two symmetric tensor irrelevant to the vector is extracted.
By means of establishing the relationship among the rank-two tensor
$\frac{\partial{L}}{\partial {g}^{\mu\nu}}$ and the rank-$(i+4)$ ones
$Z_{(i)}^{\lambda_1\cdot\cdot\cdot\lambda_i\mu\nu\rho\sigma}$s, then it
is strictly proved that the resulting rank-two symmetric tensor
is identified with the functional derivative of the Lagrangian density
with respect to the metric. This equality further gives rise to that
the field equations arising from the surface term are in agreement with
the Euler-Lagrange equations. Apart from this, it is illustrated that
the off-shell Noether current can be derived out of the surface
term as well.

First, with the surface term (\ref{SurfT}) at hand, we concentrate on the
behaviour of $\Theta^\mu$ under the transformation
$\delta\rightarrow\mathcal{L}_\zeta$,
where $\mathcal{L}_\zeta$ denotes the Lie derivative along an arbitrary
smooth vector field $\zeta^\mu$. Our main goal is to put
the surface term $\Theta^\mu(\delta\rightarrow\mathcal{L}_\zeta)$
into the form displayed by Eq. (\ref{ThetLieGF}). The detailed procedure
for doing so is presented in Appendix \ref{appendA}. So we directly
adopt the related results there. As a beginning, we
perform the transformation $\delta\rightarrow\mathcal{L}_\zeta$
to the vector $\Theta^\mu_{P}$, which leads to \cite{JJPEQM}
\be
\Theta^\mu_{P}(\delta\rightarrow\mathcal{L}_\zeta)
=2\zeta_\nu\left({P}^{\mu\lambda\rho\sigma}{R}^{\nu}_{~\lambda\rho\sigma}
-2\nabla_{\rho}\nabla_{\sigma}
{P}^{\rho\mu\nu\sigma}\right)
-\nabla_\nu{K}^{\mu\nu}_{P}
\, , \label{ThetPLie}
\ee
in which the skew-symmetric tensor ${K}^{\mu\nu}_{P}$ is in the form
\be
{K}^{\mu\nu}_{P}=2{P}^{\mu\nu\rho\sigma}
\nabla_{\rho}\zeta_{\sigma}
+4\zeta_\rho\nabla_\sigma {P}^{\mu\nu\rho\sigma}
-6{P}^{\mu[\nu\rho\sigma]}\nabla_\rho\zeta_\sigma
\, . \label{Kpmndef}
\ee
Besides, as a straightforward application of Eq. (\ref{ThetiLieGen}),
replacing the variation operator $\delta$ in
$\bar{\Theta}^\mu_{(i)}$ with the Lie derivative $\mathcal{L}_\zeta$
gives rise to
\be
\bar{\Theta}^\mu_{(i)}(\delta\rightarrow\mathcal{L}_\zeta)
=2\zeta_\nu{W}^{\mu\nu}_{(i)}
-\nabla_\nu\bar{K}^{\mu\nu}_{(i)}
\, . \label{BarThetLie}
\ee
Within Eq. (\ref{BarThetLie}), by the aid of Eqs. (\ref{Phimnidef})
and (\ref{Psimnidef}), the second-rank tensor
$W^{\mu\nu}_{(i)}=\frac{1}{2}
\big(\Phi^{\mu\nu}_{(i)}-\nabla_\lambda\Psi^{\nu\mu\lambda}_{(i)}\big)$
is expressed as
\bea
W^{\mu\nu}_{(i)}&=&\frac{1}{2}
\sum^i_{k=1}(-1)^{k-1}\Big(\nabla_{\lambda_{k-1}}\cdot\cdot\cdot
\nabla_{\lambda_{1}}Z^{\lambda_1\cdot\cdot\cdot\lambda_{k-1}\mu
\lambda_{k+1}\cdot\cdot\cdot\lambda_{i}\alpha\beta\rho\sigma}_{(i)}
\Big) \nabla^\nu \nabla_{\lambda_{k+1}}
\cdot\cdot\cdot\nabla_{\lambda_{i}}R_{\alpha\beta\rho\sigma}
 \nn \\
&&+\frac{1}{2}\nabla_\lambda\left({U}^{(\mu\nu)\lambda}_{(i)}
-{U}^{\lambda(\mu\nu)}_{(i)}
+{U}^{[\mu|\lambda|\nu]}_{(i)}\right)
\, , \label{Widef}
\eea
and the anti-symmetric rank-two tensor
$\bar{K}^{\mu\nu}_{(i)}=
-\zeta_\lambda\Psi^{\lambda\mu\nu}_{(i)}$
is presented by
\be
\bar{K}^{\mu\nu}_{(i)}=
\zeta_\lambda
\left({U}^{[\mu\nu]\lambda}_{(i)}
+{U}^{\lambda[\mu\nu]}_{(i)}
+{U}^{[\mu|\lambda|\nu]}_{(i)}\right)
\, . \label{Kbarmn}
\ee
Thus, by means of the combination of Eqs. (\ref{ThetPLie}) and
(\ref{BarThetLie}), the surface term $\Theta^\mu$ under the
transformation $\delta\rightarrow\mathcal{L}_\zeta$ turns into
the form having the same structure as the one in Eq. (\ref{ThetLieGF}),
that is,
\be
\boxed{\Theta^\mu(\delta\rightarrow\mathcal{L}_\zeta)
=2\zeta_\nu{X}^{\mu\nu}
-\nabla_\nu{Q}^{\mu\nu}}
\, . \label{ThetaLie}
\ee
In Eq. (\ref{ThetaLie}), the rank-two tensor ${X}^{\mu\nu}$
given by Eq. (\ref{GenXdef}) is recast into
\be
{X}^{\mu\nu}={P}^{\mu\lambda\rho\sigma}
{R}^{\nu}_{~\lambda\rho\sigma}-2\nabla_{\rho}\nabla_{\sigma}
{P}^{\rho\mu\nu\sigma}+{W}^{\mu\nu}
\, , \label{Xmndef}
\ee
with the ${W}^{\mu\nu}$ tensor given by
\be
{W}^{\mu\nu}=\sum^m_{i=0}W^{\mu\nu}_{(i)}
\, . \label{Wmndef}
\ee
In terms of the anti-symmetric tensors $\bar{K}^{\mu\nu}_{(i)}$s,
the anti-symmetric tensor ${Q}^{\mu\nu}$ given by Eq. (\ref{GenQdef})
can be rewritten as
\bea
{Q}^{\mu\nu}&=&2{P}^{\mu\nu\rho\sigma}
\nabla_{\rho}\zeta_{\sigma}
+4\zeta_\rho\nabla_\sigma {P}^{\mu\nu\rho\sigma}
-6{P}^{\mu[\nu\rho\sigma]}\nabla_\rho\zeta_\sigma \nn \\
&&+\zeta_\lambda\sum^m_{i=0}
\left({U}^{[\mu\nu]\lambda}_{(i)}
+{U}^{\lambda[\mu\nu]}_{(i)}
+{U}^{[\mu|\lambda|\nu]}_{(i)}\right)
\, . \label{Qmndef}
\eea
As what has been demonstrated in Appendix \ref{appendA}, in
order to guarantee that the counterpart of the surface term
$\Theta^\mu$ under $\delta\rightarrow\mathcal{L}_\zeta$ takes
on the structure displayed by Eq. (\ref{ThetaLie}), here we
stress once again that the single requirement is that there
exists an appropriate rank-two tensor ${X}^{\mu\nu}$ enabling the
difference between $\Theta^\mu(\delta\rightarrow\mathcal{L}_\zeta)$
and $2\zeta_\nu{X}^{\mu\nu}$ to be divergence-free. In view of
our straightforward calculations, the existence of ${X}^{\mu\nu}$ is
always guaranteed within the framework of the diffeomorphism invariant
gravity described by the Lagrangian (\ref{LagCovR}). This is also
supported by Eq. (\ref{LieDLtoThet}) or (\ref{RelDivThLieVLag}).
What is more, within Appendix \ref{appendA}, it is illustrated that
the rank-two tensor contracted with the arbitrary vector $\zeta^\mu$
in the first term at the right hand side of Eq. (\ref{ThetLieGF}) is
unique under the condition that the surface term with the variation
operator replaced by the Lie derivative exhibits the structure appearing
in this equation. That is to say, the tensor ${X}^{\mu\nu}$ maintains
unchanged accompanied by the alterations to the anti-symmetric tensor
${Q}^{\mu\nu}$, which is determined up to the divergence of an
arbitrary three-form.

Next, we switch to focusing on the derivation of the field equations
and conserved current out of the surface term. As what has been
demonstrated in Ref. \cite{JJPEQM}, the surface term (\ref{ThetaLie})
completely accounts for the equations of motion associated to the
Lagrangian (\ref{LagCovR}). In order to avoid confusion with
the Euler-Lagrange expression ${E}^{\mu\nu}$ given by
Eq. (\ref{EulerLagEq}), throughout this paper, the second-rank tensor
$\tilde{E}^{\mu\nu}$ is utilized to represent the expression
for such equations of motion stemming from the surface term,
and it is expressed in terms of the second-rank tensor ${X}^{\mu\nu}$
as
\be
\tilde{E}^{\mu\nu}={X}^{\mu\nu}-\frac{1}{2}{L}g^{\mu\nu}
\, . \label{EoMtild}
\ee
According to the procedure put forward in Ref. \cite{JJPEQM}, which
can be also found in Appendix \ref{appendA}, an off-shell Noether
current involving both the expression for the field equations and
the surface term with $\delta$ substituted by $\mathcal{L}_\zeta$
plays a fundamental role in producing the field equations under
the condition that the surface term
$\Theta^\mu(\delta\rightarrow\mathcal{L}_\zeta)$ possesses the
form (\ref{ThetaLie}). The variation equation (\ref{VarLag3})
of the Lagrangian under the transformation
$\delta\rightarrow\mathcal{L}_\zeta$ generates an identity
that connects the divergence of this current with that for the
expression of the Euler-Lagrange equations (see Eq. (\ref{DivJE})
or (\ref{VarLaggenLie}) for it). In essence, this
identity establishes a concrete connection between ${X}^{\mu\nu}$
and ${E}^{\mu\nu}$. Attributed to the fact that such an identity
is valid for any vector field $\zeta^\mu$, there has to exist
the one-to-one correspondence between the second-rank tensor
${X}^{\mu\nu}$ and the expression ${E}^{\mu\nu}$ for the
Euler-Lagrange equations in such a way that
${E}^{\mu\nu}={X}^{\mu\nu}-\frac{1}{2}{L}g^{\mu\nu}$, as is
shown by Eq. (\ref{Emngen}). By comparison with Eq. (\ref{EoMtild}),
then one immediately arrives at the conclusion
$\tilde{E}^{\mu\nu}={E}^{\mu\nu}$. However, for the sake of
offering a particularly straightforward proof rather than resorting
to the identity related to the off-shell Noether current to verify
that $\tilde{E}^{\mu\nu}$ indeed coincides with ${E}^{\mu\nu}$
like in Ref. \cite{JJPEQM}, in what follows, we completely abandon
the aforementioned procedure from the off-shell Noether current
perspective. As an alternative route to doing so, we directly
compare ${X}_{\mu\nu}$ with $\frac{\delta{L}}{\delta{g}^{\mu\nu}}$
to confirm that there exists no difference between both of them.
This leads to $\tilde{E}^{\mu\nu}={E}^{\mu\nu}$ in a quite
direct manner.

To achieve the goal, it is the time for us to turn to the difference
$(\Delta{E})_{\mu\nu}$ between both the tensors
$\frac{\delta{L}}{\delta{g}^{\mu\nu}}$ and ${X}_{\mu\nu}$,
namely,
\be
(\Delta{E})_{\mu\nu}
={X}_{\mu\nu}-\frac{\delta{L}}{\delta{g}^{\mu\nu}}
\, . \label{DeltEmnXL}
\ee
Alternatively, the rank-two tensor $(\Delta{E})_{\mu\nu}$ can be
defined in terms of the difference between ${E}_{\mu\nu}$ and
$\tilde{E}_{\mu\nu}$ as
\bea
(\Delta{E})_{\mu\nu}
&=&{E}_{\mu\nu}-\tilde{E}_{\mu\nu} \nn \\
&=&\frac{\partial{L}}{\partial{g}^{\mu\nu}}+
{B}_{\mu\nu}-2{P}_{\mu}^{~\lambda\rho\sigma}
{R}_{\nu\lambda\rho\sigma}
-{W}_{\mu\nu}
\, . \label{DeltEmn0}
\eea
The substitution of Eqs. (\ref{Bmndef}) and (\ref{Wmndef}) into
Eq. (\ref{DeltEmn0}) yields
\bea
(\Delta{E})^{\mu\nu}
&=&-\frac{1}{2}\sum^m_{i=0}\sum^i_{k=1}(-1)^{k-1}
\Big(\nabla_{\lambda_{k-1}}\cdot\cdot\cdot
\nabla_{\lambda_{1}}Z^{\lambda_1\cdot\cdot\cdot\lambda_{k-1}\mu
\lambda_{k+1}\cdot\cdot\cdot\lambda_{i}\alpha\beta\rho\sigma}_{(i)}
\Big) \nn \\
&&\times\nabla^\nu \nabla_{\lambda_{k+1}}
\cdot\cdot\cdot\nabla_{\lambda_{i}}R_{\alpha\beta\rho\sigma} \nn \\
&&+{g}^{\mu\rho}{g}^{\nu\sigma}
\frac{\partial{L}}{\partial{g}^{\rho\sigma}}
-2{P}^{\mu\lambda\rho\sigma}{R}^{\nu}_{~\lambda\rho\sigma}
+\frac{1}{2}\sum^m_{i=0}\nabla_\lambda{U}^{\nu\lambda\mu}_{(i)}
\, , \label{DeltEmn}
\eea
where the rank-two contravariant tensor
$(\Delta{E})^{\mu\nu}={g}^{\mu\rho}{g}^{\nu\sigma}
(\Delta{E})_{\rho\sigma}$.
For the sake of simplifying Eq. (\ref{DeltEmn}), it is useful to
introduce the following equality
\bea
\nabla_\alpha\left(\sum^{n-1}_{\ell=0}
{A}^{\nu\alpha\mu}_{(i,n;\ell)}\right)
&=&(-1)^{n}\Big(\nabla_{\lambda_{n}}\cdot\cdot\cdot
\nabla_{\lambda_{1}}Z^{\lambda_1\cdot\cdot\cdot
\lambda_{i}\mu\gamma\rho\sigma}_{(i)}\Big)\nabla_{\lambda_{n+1}}
\cdot\cdot\cdot\nabla_{\lambda_{i}}{R}^\nu_{~\gamma\rho\sigma}\nn\\
&&-Z^{\lambda_1\cdot\cdot\cdot\lambda_{i}\mu\gamma\rho\sigma}_{(i)}
\nabla_{\lambda_{1}}\cdot\cdot\cdot\nabla_{\lambda_{i}}
{R}^\nu_{~\gamma\rho\sigma}
\, , \label{IdentA}
\eea
where the third-rank tensor ${A}^\mu_{(i,n;\ell)}$
$(n=1,2,\cdot\cdot\cdot,i;
\ell=0,1,\cdot\cdot\cdot,n-1)$ is defined as
\be
{A}^{\nu\alpha\mu}_{(i,n;\ell)}=
(-1)^{n-\ell}\Big(\nabla_{\lambda_{n-\ell-1}}\cdot\cdot\cdot
\nabla_{\lambda_{1}}Z^{\lambda_1\cdot\cdot\cdot\lambda_{n-\ell-1}\alpha
\lambda_{n-\ell+1}\cdot\cdot\cdot\lambda_{i}\mu\gamma\rho\sigma}_{(i)}
\Big)\nabla_{\lambda_{n-\ell+1}}
\cdot\cdot\cdot\nabla_{\lambda_{i}}{R}^\nu_{~\gamma\rho\sigma}
\, , \label{Anamdef}
\ee
with ${A}^{\nu\alpha\mu}_{(0,n;\ell)}\equiv0$ and
${A}^{\nu\alpha\mu}_{(i,n;\ell)}={A}^{\nu\alpha\mu}_{(i,p;q)}$ under
the condition $n-\ell=p-q$. Particularly, the contraction
between the indices $\mu$ and $\nu$ in Eq. (\ref{IdentA}) produces
the equality related to the vector ${A}^{\mu}_{(i,n;\ell)}$
in \cite{PL2508}. By the aid of equality (\ref{IdentA}),
the rank-two tensor
${P}^{\mu\lambda\rho\sigma}{R}^{\nu}_{~\lambda\rho\sigma}$
can be written as
\bea
{P}^{\mu\lambda\rho\sigma}{R}^{\nu}_{~\lambda\rho\sigma}
&=&\sum^{m}_{i=0}
{Z}^{\lambda_1\cdot\cdot\cdot\lambda_{i}\mu\lambda\rho\sigma}_{(i)}
\nabla_{\lambda_{1}}\cdot\cdot\cdot\nabla_{\lambda_{i}}
{R}^{\nu}_{~\lambda\rho\sigma}
+\sum^{m}_{i=0}\sum^{i}_{k=1}
\nabla_\lambda{A}^{\nu\lambda\mu}_{(i,i;i-k)}
\, .\label{PRcontrac}
\eea
What is more, it can be verified that the divergence for the third-rank
tensor ${U}^{\nu\lambda\mu}_{(i)}$ is connected with the sum for
the divergence of ${A}^{\nu\lambda\mu}_{(i,i;i-k)}$ in the following manner
\bea
\nabla_\lambda{U}^{\nu\lambda\mu}_{(i)}&=&
\sum^i_{k=1}(-1)^{k-1}
\Big(\nabla_{\lambda_{k-1}}\cdot\cdot\cdot
\nabla_{\lambda_{1}}Z^{\lambda_1\cdot\cdot\cdot\lambda_{k-1}\mu
\lambda_{k+1}\cdot\cdot\cdot\lambda_{i}\alpha\beta\rho\sigma}_{(i)}
\Big)\nabla^\nu \nabla_{\lambda_{k+1}}
\cdot\cdot\cdot\nabla_{\lambda_{i}}R_{\alpha\beta\rho\sigma} \nn \\
&&-\sum^i_{k=1}{Z}^{\lambda_1\cdot\cdot\cdot\lambda_{k-1}\mu
\lambda_{k+1}\cdot\cdot\cdot\lambda_{i}\alpha\beta\rho\sigma}_{(i)}
\nabla_{\lambda_{1}}\cdot\cdot\cdot
\nabla_{\lambda_{k-1}}\nabla^\nu \nabla_{\lambda_{k+1}}
\cdot\cdot\cdot\nabla_{\lambda_{i}}R_{\alpha\beta\rho\sigma} \nn \\
&&+4\sum^{i}_{k=1}\nabla_\lambda{A}^{\nu\lambda\mu}_{(i,i;i-k)}
\, . \label{NabUinulmu}
\eea
Substituting Eqs. (\ref{PRcontrac}) and (\ref{NabUinulmu}) into
Eq. (\ref{DeltEmn}), one observes that the tensor
$(\Delta{E})^{\mu\nu}$ is reformulated as the form in the absence of
the arbitrary order covariant derivatives of the
${Z}^{\lambda_1\cdot\cdot\cdot\lambda_{i}\mu\nu\rho\sigma}_{(i)}$
tensor, that is,
\bea
(\Delta{E})^{\mu\nu}
&=&-\frac{1}{2}\sum^m_{i=0}\sum^i_{k=1}
{Z}^{\lambda_1\cdot\cdot\cdot\lambda_{k-1}\mu
\lambda_{k+1}\cdot\cdot\cdot\lambda_{i}\alpha\beta\rho\sigma}_{(i)}
\nabla_{\lambda_{1}}\cdot\cdot\cdot
\nabla_{\lambda_{k-1}}\nabla^\nu \nabla_{\lambda_{k+1}}
\cdot\cdot\cdot\nabla_{\lambda_{i}}R_{\alpha\beta\rho\sigma} \nn \\
&&-2\sum^{m}_{i=0}
{Z}^{\lambda_1\cdot\cdot\cdot\lambda_{i}\mu\lambda\rho\sigma}_{(i)}
\nabla_{\lambda_{1}}\cdot\cdot\cdot\nabla_{\lambda_{i}}
{R}^{\nu}_{~\lambda\rho\sigma}
+{g}^{\mu\rho}{g}^{\nu\sigma}
\frac{\partial{L}}{\partial{g}^{\rho\sigma}}
\, . \label{DeltEmn2}
\eea
By the comparison between Eqs. (\ref{DeltEmn0}) and (\ref{DeltEmn2}),
we have proved by performing direct computation that there exists
the identity expressing the linear combination of the three tensors
${W}^{\mu\nu}$, ${B}^{\mu\nu}$ and ${P}^{\mu\lambda\rho\sigma}
{R}^{\nu}_{~\lambda\rho\sigma}$ in the following manner
\bea
{W}^{\mu\nu}&=&{B}^{\mu\nu}-2{P}^{\mu\lambda\rho\sigma}
{R}^{\nu}_{~\lambda\rho\sigma}
+2\sum^{m}_{i=0}
{Z}^{\lambda_1\cdot\cdot\cdot\lambda_{i}\mu\lambda\rho\sigma}_{(i)}
\nabla_{\lambda_{1}}\cdot\cdot\cdot\nabla_{\lambda_{i}}
{R}^{\nu}_{~\lambda\rho\sigma} \nn \\
&&+\frac{1}{2}\sum^m_{i=0}\sum^i_{k=1}
{Z}^{\lambda_1\cdot\cdot\cdot\lambda_{k-1}\mu
\lambda_{k+1}\cdot\cdot\cdot\lambda_{i}\alpha\beta\rho\sigma}_{(i)}
\nabla_{\lambda_{1}}\cdot\cdot\cdot
\nabla_{\lambda_{k-1}}\nabla^\nu \nabla_{\lambda_{k+1}}
\cdot\cdot\cdot\nabla_{\lambda_{i}}R_{\alpha\beta\rho\sigma}
\, . \label{PWBexpan}
\eea
Besides, it will be shown by Eqs. (\ref{ZPDivWLie2}) and
(\ref{Upsimndef}) below that the identity (\ref{PWBexpan}) can
be alternatively derived out of Eq. (\ref{ZidelnabR}) with the
substitution $\delta\rightarrow\mathcal{L}_\zeta$.

Note that Eq. (\ref{DeltEmn2}) is not the simplest form that
$(\Delta{E})^{\mu\nu}$ actually belongs to. As a matter of fact,
one is able to go further. To see this, like in the work
\cite{Pady}, it is feasible to resort to the Lie derivative
of the Lagrangian density ${L}$ along the vector field
$\zeta^\mu$. In practice, this is able to be obtained by means of the
setting that the variation of the Lagrangian density ${L}$ is
performed via the Lie derivative $\mathcal{L}_\zeta$ acting on
${L}$. Accordingly, replacing the $\delta$ operator in the variation
equation (\ref{VariLagdens}) of the Lagrangian density with
$\mathcal{L}_\zeta$ gives rise to
\be
\boxed{\mathcal{L}_\zeta{L}=
\frac{\partial{L}}{\partial{g}^{\mu\nu}}\mathcal{L}_\zeta{g}^{\mu\nu}
+\sum^{m}_{i=0}
\frac{\partial{L}}{\partial\nabla_{\lambda_1}\cdot\cdot\cdot
\nabla_{\lambda_i}{R}_{\mu\nu\rho\sigma}}
\mathcal{L}_\zeta\nabla_{\lambda_1}
\cdot\cdot\cdot\nabla_{\lambda_i}R_{\mu\nu\rho\sigma}}
\, . \label{LieLagdens}
\ee
That is to say, equation (\ref{LieLagdens}) can be understood as an
outcome arising from the variation of the Lagrangian density, or that
for the Lagrangian in the manner $\mathcal{L}_\zeta\mathcal{L}
=\sqrt{-g}(\mathcal{L}_\zeta{L}+{L}\nabla_\mu\zeta^\mu)$.
Here we point out that Eq. (\ref{LieLagdens}) is sufficient for
our purpose. It is unnecessary for us to introduce the other expression
for $\mathcal{L}_\zeta{L}$ that consists of the covariant derivative
of the Lagrangian density ${L}$ (see Eq. (\ref{CovdeL}) below for its
explicit expression) as shown in Ref. \cite{Pady}. According to the
rules of Lie derivative acting on tensors, we observe that equation
(\ref{LieLagdens}) can be further recast into
\be
\zeta^\nu\left(\nabla_\nu{L}
-\sum^{m}_{i=0}
Z_{(i)}^{\lambda_1\cdot\cdot\cdot\lambda_i\alpha\beta\rho\sigma}
\nabla_\nu\nabla_{\lambda_1}
\cdot\cdot\cdot\nabla_{\lambda_i}R_{\alpha\beta\rho\sigma}\right)
+2(\Delta{E})^{\mu\nu}\nabla_\mu\zeta_\nu=0
\, . \label{LieLagdens2}
\ee
The above equation holds for any vector field $\zeta^\mu$.
As a result of this, there have to exist the following two identities,
given respectively by
\bea
\nabla_\nu{L}&=&\sum^{m}_{i=0}
Z_{(i)}^{\lambda_1\cdot\cdot\cdot\lambda_i\alpha\beta\rho\sigma}
\nabla_\nu\nabla_{\lambda_1}
\cdot\cdot\cdot\nabla_{\lambda_i}{R}_{\alpha\beta\rho\sigma}
\, , \label{CovdeL} \\
(\Delta{E})^{\mu\nu}&=&0
\, . \label{VanishDelE}
\eea
Equation (\ref{CovdeL}) stands for the expression for the covariant
derivative of the Lagrangian density expanded in the variables
$\nabla_{\lambda_1}\cdot\cdot\cdot\nabla_{\lambda_i}
{R}_{\alpha\beta\rho\sigma}$s $(i=0,1,\cdot\cdot\cdot,m)$. Unlike
in Ref. \cite{Pady}, we here derive it out of the Lie derivative of
the Lagrangian density rather than treat it as the other condition to
produce the identity (\ref{VanishDelE}) in addition to the one given
by Eq. (\ref{LieLagdens}). That is to say, within the context of
Padmanabhan's method, equation (\ref{LieLagdens}) is actually the
sufficient condition for the derivation of the identity (\ref{VanishDelE}),
which is adequately responsible for the elimination of the derivative of
the Lagrangian density with respect to the metric appearing in the
field equations. Moreover, as shown in Eq. (\ref{DivXmn2}), the identity
(\ref{CovdeL}) will play an important role in proving the vanishing of the
divergence for the expression of the field equations by carrying out
direct computation.

From Eq. (\ref{VanishDelE}), we directly obtain the following identity
\bea
{g}^{\mu\rho}{g}^{\nu\sigma}
\frac{\partial{L}}{\partial{g}^{\rho\sigma}}&=&
\frac{1}{2}\sum^m_{i=0}\sum^i_{k=1}
{Z}^{\lambda_1\cdot\cdot\cdot\lambda_{k-1}\mu
\lambda_{k+1}\cdot\cdot\cdot\lambda_{i}\alpha\beta\rho\sigma}_{(i)}
\nabla_{\lambda_{1}}\cdot\cdot\cdot
\nabla_{\lambda_{k-1}}\nabla^\nu \nabla_{\lambda_{k+1}}
\cdot\cdot\cdot\nabla_{\lambda_{i}}R_{\alpha\beta\rho\sigma} \nn \\
&&+2\sum^{m}_{i=0}
{Z}^{\lambda_1\cdot\cdot\cdot\lambda_{i}\mu\lambda\rho\sigma}_{(i)}
\nabla_{\lambda_{1}}\cdot\cdot\cdot\nabla_{\lambda_{i}}
{R}^{\nu}_{~\lambda\rho\sigma}\nn \\
&=&2{P}^{\mu\lambda\rho\sigma}
{R}^{\nu}_{~\lambda\rho\sigma}
+{W}^{\mu\nu}-{B}^{\mu\nu}
\, . \label{Partgmnequal}
\eea
The above identity associates the derivative of the Lagrangian density
with respect to the metric with the one with regard to the covariant
derivatives of the Riemann tensor. As is what will be demonstrated in
Eq. (\ref{ZPDivWLie3}) below, the equality (\ref{Partgmnequal}) can be
reproduced in an alternative way. Apart from Eq. (\ref{Partgmnequal}),
owing to the fact that the second-rank tensors ${B}^{\mu\nu}$ and
$\frac{\partial{L}}{\partial{g}^{\mu\nu}}$ are symmetric,
$(\Delta{E})^{[\mu\nu]}=0$ appearing as a result from Eq. (\ref{VanishDelE})
leads to another identity
\be
{W}^{[\mu\nu]}=-2{P}^{[\mu|\lambda\rho\sigma|}
{R}^{\nu]}_{~~\lambda\rho\sigma}
\, , \label{WPantiequal}
\ee
or equivalently,
\bea
&&\sum^m_{i=0}\sum^i_{k=1}
{Z}^{\lambda_1\cdot\cdot\cdot\lambda_{k-1}[\mu
|\lambda_{k+1}\cdot\cdot\cdot\lambda_{i}\alpha\beta\rho\sigma|}_{(i)}
\nabla_{\lambda_{1}}\cdot\cdot\cdot
\nabla_{\lambda_{k-1}}\nabla^{\nu]}\nabla_{\lambda_{k+1}}
\cdot\cdot\cdot\nabla_{\lambda_{i}}R_{\alpha\beta\rho\sigma} \nn \\
&&+4\sum^{m}_{i=0}
{Z}^{\lambda_1\cdot\cdot\cdot\lambda_{i}[\mu|\lambda\rho\sigma|}_{(i)}
\nabla_{\lambda_{1}}\cdot\cdot\cdot\nabla_{\lambda_{i}}
{R}^{\nu]}_{~~\lambda\rho\sigma}=0
\, . \label{WPantiequal2}
\eea
Equation (\ref{WPantiequal}) or (\ref{WPantiequal2}) reveals
that the second-rank tensor ${X}^{\mu\nu}$ is symmetric,
resulting in $\tilde{E}^{\mu\nu}=\tilde{E}^{\nu\mu}$.
In particular, when the Lagrangian (\ref{LagCovR}) depends on
${g}^{\mu\nu}$ and $\Box^i{R}_{\mu\nu\rho\sigma}$s, both the
two identities (\ref{Partgmnequal}) and (\ref{WPantiequal2})
coincide with the corresponding results given by \cite{PL2402}. What is
more, as a direct consequence of Eq. (\ref{VanishDelE}), we have
proved strictly that
\be
{X}_{\mu\nu}=\frac{\delta{L}}{\delta{g}^{\mu\nu}}
\, . \label{XisdelLd}
\ee
Equation (\ref{XisdelLd}) yields in a straightforward way
that both ${E}^{\mu\nu}$ and $\tilde{E}^{\mu\nu}$ are completely
identified with each other, that is,
\be
{E}^{\mu\nu}=\tilde{E}^{\mu\nu}={P}^{\mu\lambda\rho\sigma}
{R}^{\nu}_{~\lambda\rho\sigma}-2\nabla_{\rho}\nabla_{\sigma}
{P}^{\rho\mu\nu\sigma}-\frac{1}{2}{L}g^{\mu\nu}+{W}^{\mu\nu}
\, . \label{EequltldE}
\ee
This implies that we verify once again the consequence
${E}^{\mu\nu}=\tilde{E}^{\mu\nu}$ obtained in our previous work
\cite{JJPEQM} in a different way. From equation (\ref{EequltldE}),
it is indicated that it is unnecessary to distinguish ${E}^{\mu\nu}$
and $\tilde{E}^{\mu\nu}$ in essence, while the tilde over the
expression for the field equations only represents that this
expression is derived out of the surface term. Nonetheless, one
observes that employing ${X}_{\mu\nu}$ as a replacement for
$\frac{\delta{L}}{\delta{g}^{\mu\nu}}$ completely avoids the
calculation on Eq. (\ref{Partgmnequal}) and this makes
$\tilde{E}^{\mu\nu}$ naturally eliminate the tensor
$\frac{\partial{L}}{\partial{g}^{\mu\nu}}$ so that
$\tilde{E}^{\mu\nu}$ takes a simpler form than ${E}^{\mu\nu}$
given by Eq. (\ref{EulerLagEq}). Besides, the composition manner
of $\tilde{E}^{\mu\nu}$ provides convenience for verifying the
vanishing for the covariant divergence of the expression of
the field equations via straightforward computations as is what
will be demonstrated in Section \ref{six}. Moreover, some
properties on the trace of $\tilde{E}^{\mu\nu}$ can be found
in \cite{PL2508}.

In the above, it has been proved that the expression $\tilde{E}^{\mu\nu}$
for the field equations appears as a consequence from the surface term.
However, due to the fact that equation (\ref{Partgmnequal}) allows for the
removal of the tensor $\frac{\partial{L}}{\partial{g}^{\mu\nu}}$
or $\frac{\partial{L}}{\partial{g}_{\mu\nu}}$ in the field equations,
$\tilde{E}_{\mu\nu}$ is also able to be thought of as the result arising
from the substitution of Eq. (\ref{Partgmnequal}) into the Euler-Lagrange
expression ${E}_{\mu\nu}$ given by Eq. (\ref{EulerLagEq}), that is,
\be
\tilde{E}_{\mu\nu}={E}_{\mu\nu}
\left(\frac{\partial{L}}{\partial{g}^{\mu\nu}}\rightarrow
2{P}_\mu^{~\lambda\rho\sigma}{R}_{\nu\lambda\rho\sigma}
+{W}_{\mu\nu}-{B}_{\mu\nu}\right)
\, . \label{EoMPady}
\ee
In fact, this is fully in accordance with the spirit of the procedure
proposed by Padmanabhan to derive the equations of motion associated
to the theories of gravity with the Lagrangian only built out of the
metric and its Riemann tensor within Ref. \cite{Pady} (the key point
of Padmanabhan's method lies within Eq. (\ref{EquiReLieLE}) or the
first equivalence relation of Eq. (\ref{LieLEJequiR}) in Section
\ref{five}). To this point, $\tilde{E}^{\mu\nu}$ can be regarded as
the most natural generalization for the expression of the equations
of motion given by Ref. \cite{Pady}
to the theories of gravity involving the arbitrary order covariant
derivatives of the Riemann tensor. Consequently, just from the
perspective for the derivation of the equations of motion, our procedure
for the extraction of the field equations out of the surface term
provides another path towards the reproduction and the generalization
for the field equations via Padmanabhan's method, while the former
offers an additional advantage of ruling out the calculations involving
the derivative of the Lagrangian density with respect to the metric
tensor.

Finally, apart from the equations of motion for the gravitational
field, we demonstrate that the surface term (\ref{ThetaLie}) also
encodes the information for conserved current. To see this, we turn
our attention to extract the conserved current from such a term.
Substituting $X^{\mu\nu}={E}^{\mu\nu}+\frac{1}{2}{g}^{\mu\nu}L$ into
Eq. (\ref{ThetaLie}), we obtain the conserved current ${J}^\mu$
associated to the vector field $\zeta^\mu$, defined in terms of
the quantity $\Theta^\mu(\delta\rightarrow\mathcal{L}_\zeta)$ through
\be
{J}^\mu=2\zeta_\nu{E}^{\mu\nu}+\zeta^\mu{L}
-\Theta^\mu(\delta\rightarrow\mathcal{L}_\zeta)
\, , \label{ConsCurr}
\ee
or equivalently defined in terms of the anti-symmetric tensor
${Q}^{\mu\nu}$ as ${J}^\mu=\nabla_\nu{Q}^{\mu\nu}$. Actually, here
the current ${J}^\mu$ is nothing else but the conventional off-shell
Noether current relative to the vector field $\zeta^\mu$
\cite{GJS12,TPad10,RievLL}, while the anti-symmetric rank-two
tensor ${Q}^{\mu\nu}$ is the corresponding Noether charge 2-form
\cite{IyerWald,NoeChWald}. We here present the explicit form
for the Noether charge 2-form within the context of the diffeomorphism
invariant theories of gravity relying on the covariant derivatives
of the Riemann tensor. This could be helpful to understand the Clausius
relation relative to the diffeomorphism invariant gravities encompassing
higher derivative and higher curvature terms \cite{DLM16,GJS12}.

\section{The comparison between two ways of identifying
the expressions for the field equations}\label{four}

We have illustrated in great detail that the expression for the
field equations can be derived out of the surface term arising
from the variation of the Lagrangian in the previous section,
as well as in our previous work \cite{JJPEQM}. The main process
to do so in both the works is the same, which consists of
three steps:
\be
\text{Evaluating }\Theta^\mu\quad\Rightarrow\quad
\text{Putting }\Theta^\mu \text{ into form } (\ref{ThetaLie})
\quad\Rightarrow\quad
\text{Confirming }{E}^{\mu\nu}=\tilde{E}^{\mu\nu}
\, . \label{ProcST}
\ee
The only difference appearing in the last step is the manner in
which we verify that both the expressions $\tilde{E}^{\mu\nu}$
and ${E}^{\mu\nu}$ for the field equations indeed coincide with
each other. Owing to this, within the present section, we will offer
a detailed comparison between the methods to identify
$\tilde{E}^{\mu\nu}$ with ${E}^{\mu\nu}$
employed respectively in the present work and Ref. \cite{JJPEQM}.
Meanwhile, motivated by the results, we will reveal several
connections between the expression for the field equations and the
off-shell Noether current.

In the previous section, it has been demonstrated that the relationship
$\tilde{E}^{\mu\nu}={E}^{\mu\nu}$ directly results
from Eq. (\ref{XisdelLd}), which arises as a consequence of the
pivotal relationship among the symmetric rank-two tensor
$\frac{\partial{L}}{\partial{g}^{\mu\nu}}$ and the derivative of
the Lagrangian density with respect to the $i$th-order covariant
derivatives of the Riemann tensor
$\frac{\partial{L}}{\partial\nabla_{\lambda_1}\cdot\cdot\cdot
\nabla_{\lambda_i}R_{\mu\nu\rho\sigma}}$s, given by
Eq. (\ref{Partgmnequal}). This equation, resulting from Eq.
(\ref{LieLagdens}), actually serves as the starting
point of identifying $\tilde{E}^{\mu\nu}$ with ${E}^{\mu\nu}$, and it
provides a solid foundation for establishing the connection
$\tilde{E}^{\mu\nu}={E}^{\mu\nu}$ since it arises as a
straightforward outcome for the Lie derivative of the Lagrangian
density along an arbitrary smooth vector field without any
other conditions. At last, the off-shell Noether current ${J}^\mu$
given by Eq. (\ref{ConsCurr}) appears as a result of
$\tilde{E}^{\mu\nu}={E}^{\mu\nu}$. To sum up, the aforementioned
process from the perspective of computing directly the difference
between ${E}^{\mu\nu}$ and $\tilde{E}^{\mu\nu}$ goes as follows:
\be
\text{Equation }(\ref{Partgmnequal})\Rightarrow
{X}_{\mu\nu}=\frac{\delta{L}}{\delta{g}^{\mu\nu}}\Rightarrow
\tilde{E}^{\mu\nu}={E}^{\mu\nu}\Rightarrow
\nabla_\mu{J}^\mu=\nabla_\mu\nabla_\nu{Q}^{\mu\nu}=0
\, . \label{ProceOne}
\ee

Apart from the above process for the handling of
$\tilde{E}^{\mu\nu}={E}^{\mu\nu}$, it has been illustrated in
our previous work \cite{JJPEQM} that such an equality
can be obtained from the off-shell Noether current perspective.
To do so, conversely, the off-shell Noether current ${J}^\mu$
is adopted to produce the expression for the field equations
rather than arises as the outcome of this expression. As is 
shown by Eq. (\ref{VarLaggenLie}), the covariant divergence
of this current is associated to the identity
\be
\boxed{\nabla_\mu{J}^\mu=2\zeta_\nu\nabla_\mu{E}^{\mu\nu}}
\, , \label{DivJE}
\ee
directly originating from the variation equation (\ref{VarLag3}) of
the Lagrangian under the transformation
$\delta\rightarrow\mathcal{L}_\zeta$. As a matter of fact, the 
identity (\ref{DivJE}), which can be viewed as a bridge
between the divergence for the conserved current and that for
the expression of the field equations, constitutes the starting
point to yield $\tilde{E}^{\mu\nu}={E}^{\mu\nu}$ in the work 
\cite{JJPEQM}. By contrast, employing the identity (\ref{DivJE}) 
instead of Eq. (\ref{Partgmnequal}) as the starting point is 
relatively simple. Besides, in that work, equation (\ref{Partgmnequal}) 
is treated as the consequence resulting from 
$\tilde{E}^{\mu\nu}={E}^{\mu\nu}$ rather than the foundation to 
verify this equality like in the present work. In short, the process 
to obtain $\tilde{E}^{\mu\nu}={E}^{\mu\nu}$ from the off-shell 
Noether current perspective can be summarized as
\be
\nabla_\mu{J}^\mu=2\zeta_\nu\nabla_\mu{E}^{\mu\nu}
\Rightarrow
\tilde{E}^{\mu\nu}={E}^{\mu\nu}\Rightarrow
{X}_{\mu\nu}=\frac{\delta{L}}{\delta{g}^{\mu\nu}}
\Rightarrow
\text{Equation }(\ref{Partgmnequal})
\, , \label{ProceTwo}
\ee
which can be regarded as the inverse of Eq. (\ref{ProceOne}).

Actually, according to Eq. (\ref{NSconTheLie}), the form (\ref{ThetaLie})
for the surface term $\Theta^\mu(\delta\rightarrow\mathcal{L}_\zeta)$
results in $\nabla_\mu{E}^{\mu\nu}=0$ under the condition (\ref{DivJE}).
That is to say, equation (\ref{DivJE}) can be replaced with the conserved
equation $\nabla_\mu{J}^\mu=0$ for the Noether current ${J}^\mu$.
Therefore, for the Lagrangian (\ref{LagCovR}) accommodating diffeomorphism
invariance, on the basis of Eqs. (\ref{ProceOne}) and (\ref{ProceTwo}),
there exist the following equivalence relations:
\be
\boxed{\text{Equation }(\ref{Partgmnequal})\quad\Leftrightarrow\quad
{X}_{\mu\nu}=\frac{\delta{L}}{\delta{g}^{\mu\nu}}\quad\Leftrightarrow\quad
{E}^{\mu\nu}=\tilde{E}^{\mu\nu}\quad\Leftrightarrow\quad
\nabla_\mu{J}^\mu=0}
\, . \label{FourEquiv}
\ee
From Eq. (\ref{FourEquiv}), we observe that both the two ways to verify
${E}^{\mu\nu}=\tilde{E}^{\mu\nu}$ presented respectively by this work
and our previous one \cite{JJPEQM} are equivalent.

\begin{figure}[h]
\centering
\includegraphics[width=12cm]{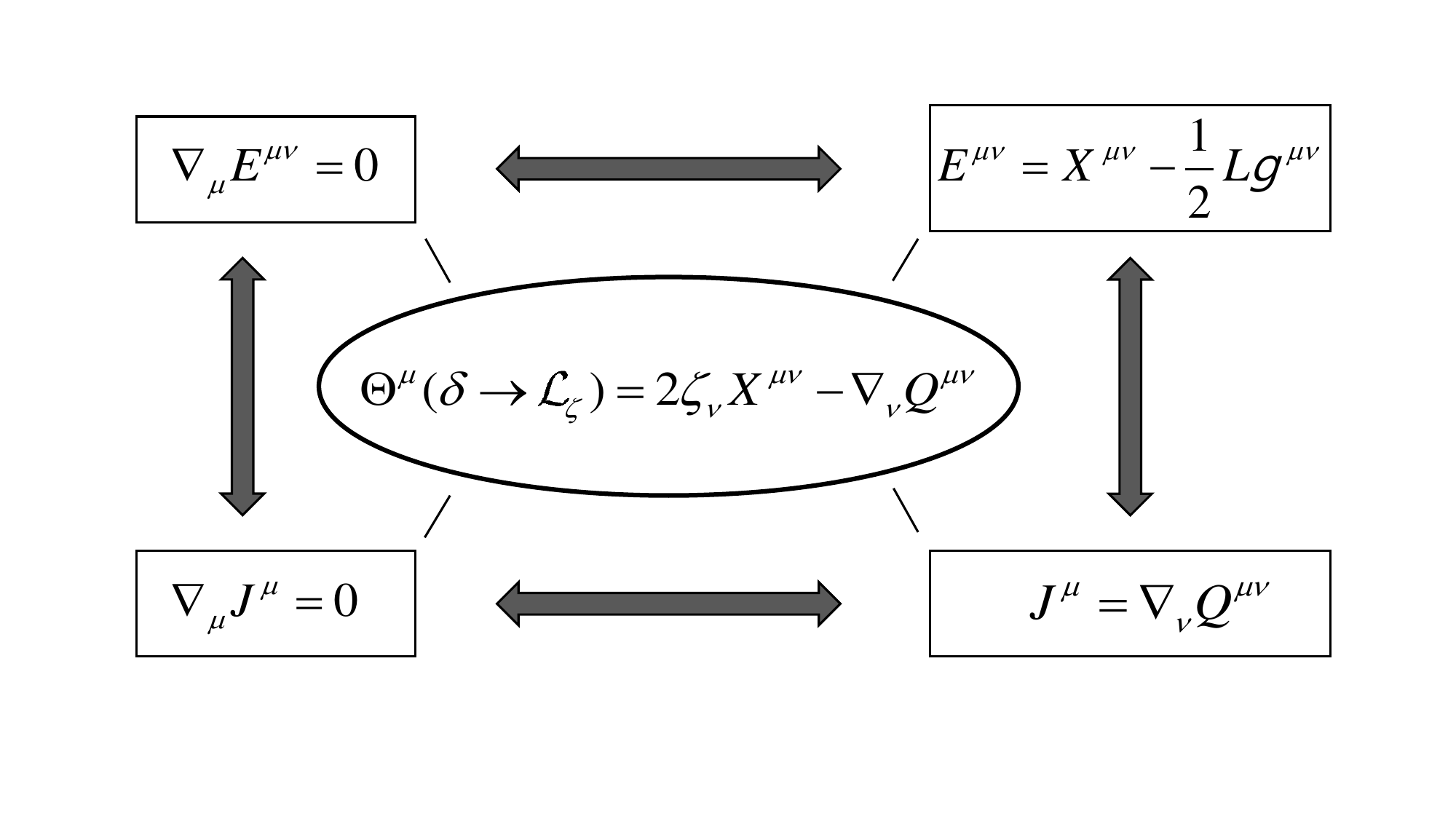}
\vspace{-1.0cm}
\caption{The equivalence relationships associated to the expression
${E}^{\mu\nu}$ for the field equations and the conserved current
${J}^\mu$ are unveiled. The definition (\ref{ConsCurr}) for the
conserved current is employed. The expression (\ref{ThetaLie}) 
for the surface term is instrumental in establishing these 
equivalence relationships. Additionally, equation (\ref{DivJE}),
which is equivalent to equation (\ref{LieLagdens}), gets involved in 
establishing the equivalence between $\nabla_\mu{E}^{\mu\nu}=0$ 
and $\nabla_\mu{J}^{\mu}=0$.}
\label{FigEJ}
\end{figure}

Furthermore, we concentrate on the relations between the conserved
Noether current and the expression for the field equations. With the
help of the correspondence
\be
\nabla_\mu{J}^\mu=0\quad\Leftrightarrow\quad
\nabla_\mu{E}^{\mu\nu}=0
\, \label{EJequivR}
\ee
arising from the identity (\ref{DivJE}), in terms of Eq. (\ref{FourEquiv}),
one further builds the equivalence relationships among the generalized
Bianchi identity $\nabla_\mu{E}^{\mu\nu}=0$, the conserved equation
$\nabla_\mu{J}^\mu=0$, the expression for the field equations and
the off-shell Noether current, given by
\be
\boxed{\nabla_\mu{J}^\mu=0\quad\Leftrightarrow\quad
\nabla_\mu{E}^{\mu\nu}=0\quad\Leftrightarrow\quad
{E}^{\mu\nu}={X}^{\mu\nu}-\frac{1}{2}{L}{g}^{\mu\nu}
\quad\Leftrightarrow\quad
{J}^\mu=\nabla_\nu{Q}^{\mu\nu}}
\, . \label{EmnequlCC}
\ee
All the equivalence relationships appearing in Eq. (\ref{EmnequlCC}) are
displayed by Fig. \ref{FigEJ}. That is to say, the expression for the
equations of motion can arise as the result of the conserved equation
for the current. Apart from this, the equivalence between
$\nabla_\mu{E}^{\mu\nu}=0$
and ${E}^{\mu\nu}={X}^{\mu\nu}-\frac{1}{2}{L}{g}^{\mu\nu}$
in Eq. (\ref{EmnequlCC}) allows for the derivation
for the expression of the field equations out of the vanishing of
its divergence. In light of this, the equivalence relationships given
by Eq. (\ref{EmnequlCC}) might provide some insights on the
thermodynamic derivation for the gravitational field equations
\cite{DLM16,GJS12,Jacob95}. Particularly, we propose that the
expression for the field equations stemming from the thermodynamic
derivation should be $\tilde{E}^{\mu\nu}$ ranther than ${E}^{\mu\nu}$.
It is worth mentioning that all the three
equations (\ref{ThetaLie}), (\ref{LieLagdens}) and (\ref{DivJE}) play a
crucial role in establishing the equivalence relationships appearing
in Eqs. (\ref{FourEquiv}) and (\ref{EmnequlCC}). In fact, as is what
will be illustrated in detail within the next section, since equation
(\ref{ThetaLie}) holds identically, all the three equations constantly
coexist as long as either of the two equations (\ref{LieLagdens})
and (\ref{DivJE}) among them comes into existence.

At the end, we observe that equation (\ref{EmnequlCC}) gives rise
to another interesting relation between the conserved current and
the expression for the field equations, that is, the structure for
the equivalence relationship associated to the off-shell Noether
current
\be
\nabla_\mu{J}^\mu=0\quad\Leftrightarrow\quad
{J}^\mu=\nabla_\nu{Q}^{\mu\nu}
\,  \label{CoCurequiv}
\ee
bears a resemblance to the one involving the expression
for the field equations
\be
\nabla_\mu{E}^{\mu\nu}=0\quad\Leftrightarrow\quad
{E}^{\mu\nu}={X}^{\mu\nu}-\frac{1}{2}{L}{g}^{\mu\nu}
\, . \label{EoMequiv}
\ee
Both the equations (\ref{CoCurequiv}) and (\ref{EoMequiv})
establish the equivalence between the concrete expression
of a quantity and the vanishing of its covariant divergence.
In light of Eqs. (\ref{EJequivR}), (\ref{CoCurequiv}) and
(\ref{EoMequiv}), together with
\be
{E}^{\mu\nu}={X}^{\mu\nu}-\frac{1}{2}{L}{g}^{\mu\nu}
\quad\Leftrightarrow\quad
{J}^\mu=\nabla_\nu{Q}^{\mu\nu}
\, , \label{EoMCCequiv}
\ee
we propose that there exists the ${E}^{\mu\nu}/{J}^\mu$
correspondence between the expression ${E}^{\mu\nu}$ for the equations
of motion and the conserved current ${J}^\mu$ in the framework of
diffeomorphism invariant theories of pure gravity.

\section{Relations among the three equations
(\ref{ThetaLie}), (\ref{LieLagdens}) and (\ref{DivJE})}
\label{five}

Due to the analysis performed in the previous two sections, the
combination of the two equations (\ref{ThetaLie}) and (\ref{LieLagdens})
plays a critical role in the derivation for the field equations without
the derivative of the Lagrangian density with respect to the metric in
the present work, while it was illustrated in our previous work
\cite{JJPEQM} that the expression for the equations of motion can
also appear as a consequence from the combination of Eqs. (\ref{ThetaLie})
and (\ref{DivJE}), that is,
\be
\text{Eqs. }(\ref{ThetaLie})\text{ \& }(\ref{LieLagdens})
\quad\Rightarrow\quad
\boxed{{E}^{\mu\nu}={X}^{\mu\nu}-\frac{1}{2}{L}{g}^{\mu\nu}}
\quad\Leftarrow\quad
\text{Eqs. }(\ref{ThetaLie})\text{ \& }(\ref{DivJE})
\, . \label{TwowaystoE}
\ee
Apparently, the three equations (\ref{ThetaLie}),
(\ref{LieLagdens}) and (\ref{DivJE}) that arise from the
substitution of the variation operator by the Lie derivative
along an arbitrary smooth vector field occupy an important
position in our study. Thus, within this section, we will reveal
some interesting and significant relations among them. Inspired
with the results, we will discuss the main implications of
Padmanabhan's method.

Our fundamental purpose for resorting to the Lie derivative of the
Lagrangian density ${L}$ along the diffeomorphism generator
$\zeta^\mu$ given by Eq. (\ref{LieLagdens}) is to establish the
connection between $\frac{\partial{L}}{\partial{g}^{\mu\nu}}$ and
$\frac{\partial{L}}{\partial\nabla_{\lambda_1}\cdot\cdot\cdot
\nabla_{\lambda_i}{R}_{\mu\nu\rho\sigma}}$s, by means of which the
rank-two tensor $\frac{\partial{L}}{\partial{g}^{\mu\nu}}$ in the
Euler-Lagrange equations can be eliminated. It has been demonstrated
that Eq. (\ref{LieLagdens}) gives rise to Eq. (\ref{Partgmnequal})
that indeed assists us to achieve this goal. In fact, equation
(\ref{LieLagdens}) is completely equivalent to the combination of
both the equations (\ref{CovdeL}) and (\ref{Partgmnequal}).
Moreover, by the aid of Eq. (\ref{DivXmn2}), equation (\ref{CovdeL})
is able to arise as a consequence from the generalized Bianchi
identity $\nabla_\mu{E}^{\mu\nu}=0$ written as
$\nabla^\nu{L}=2\nabla_\mu{X}^{\mu\nu}$ and vice versa, while
Eq. (\ref{FourEquiv}) shows that Eq. (\ref{Partgmnequal}) is
equivalent to ${E}^{\mu\nu}=\tilde{E}^{\mu\nu}$. Consequently,
we obtain
\be
\text{Equation }(\ref{LieLagdens})\quad\Leftrightarrow\quad
\text{Eqs. }(\ref{CovdeL})\text{ and }(\ref{Partgmnequal})
\quad\Leftrightarrow\quad
{E}^{\mu\nu}=\tilde{E}^{\mu\nu} \text{ and }
\nabla_\mu{E}^{\mu\nu}=0
\, . \label{EquiReLieLE}
\ee
As shown by Eq. (\ref{EquiReLieLE}), there exist two ways to
verify Eq. (\ref{LieLagdens}). One is to confirm that both
the two equations (\ref{CovdeL}) and  (\ref{Partgmnequal}) hold
through direct calculations, and the other is to test
${E}^{\mu\nu}={X}^{\mu\nu}-\frac{1}{2}{g}^{\mu\nu}{L}$ together
with the generalized Bianchi identity for ${E}^{\mu\nu}$.

What is more, according to Eq. (\ref{VarLaggenLieXQ}), under the
condition that the surface term
$\Theta^\mu(\delta\rightarrow\mathcal{L}_\zeta)$
possesses the form (\ref{ThetaLie}), the identity (\ref{DivJE})
gives rise to $\nabla_\mu{E}^{\mu\nu}=0$ and
${E}^{\mu\nu}=\tilde{E}^{\mu\nu}$ simultaneously. The former
leads to Eq. (\ref{CovdeL}), and the latter yields
Eq. (\ref{Partgmnequal}) in terms of Eq. (\ref{ProceTwo}).
Ultimately, both of Eqs. (\ref{CovdeL}) and (\ref{Partgmnequal})
jointly produce Eq. (\ref{LieLagdens}). On the other hand, by
virtue of Eq. (\ref{EquiReLieLE}), it is easy to see that
${E}^{\mu\nu}=\tilde{E}^{\mu\nu}$ together with
$\nabla_\mu{E}^{\mu\nu}=0$ can be derived out of
Eq. (\ref{LieLagdens}). Then the combination of
${E}^{\mu\nu}=\tilde{E}^{\mu\nu}$ and $\nabla_\mu{E}^{\mu\nu}=0$
further leads to Eq. (\ref{DivJE}) in the form
\bea
\nabla_\mu\big(\zeta^\mu{L}\big)&=&
-2{E}^{\mu\nu}\nabla_\mu\zeta_\nu
+\nabla_\mu\big(2\zeta_\nu{X}^{\mu\nu}\big) \nn \\
&=&-2{E}^{\mu\nu}\nabla_\mu\zeta_\nu
+\nabla_\mu\Theta^\mu(\delta\rightarrow\mathcal{L}_\zeta)
\, . \label{VaLtoLieL}
\eea
Equation (\ref{VaLtoLieL}) is just Eq. (\ref{VarLag3}) with $\delta$
substituted by $\mathcal{L}_\zeta$. In order to arrive at the second
equality in the above equation, here we have utilized the relation
$\nabla_\mu\big(2\zeta_\nu{X}^{\mu\nu}\big)
=\nabla_\mu\Theta^\mu(\delta\rightarrow\mathcal{L}_\zeta)$ resulting
from Eq. (\ref{ThetaLie}). Hence, it is illustrated that
Eq. (\ref{DivJE}) can be reproduced by Eq. (\ref{LieLagdens}) with
the help of Eq. (\ref{ThetaLie}). Due to the aforementioned analysis,
as long as $\Theta^\mu(\delta\rightarrow\mathcal{L}_\zeta)$ is
provided with the form (\ref{ThetaLie}) or its covariant divergence
coincides with that for the vector $2\zeta_\nu{X}^{\mu\nu}$, we have
\be
\boxed{\text{Equation }(\ref{LieLagdens})\quad\Leftrightarrow\quad
{E}^{\mu\nu}=\tilde{E}^{\mu\nu} \text{ and }
\nabla_\mu{E}^{\mu\nu}=0
\quad\Leftrightarrow\quad
\text{Equation }(\ref{DivJE})}
\, . \label{LieLEJequiR}
\ee
It is worth pointing out that the first equivalence relationship
in Eq. (\ref{LieLEJequiR}) holds without the requirement to impose
the condition that $\Theta^\mu(\delta\rightarrow\mathcal{L}_\zeta)$
admits the form (\ref{ThetaLie}), which only plays a role in the
establishment of the second equivalence relationship. As a matter
of fact, equation (\ref{EquiReLieLE}) or the first equivalence
relationship in Eq. (\ref{LieLEJequiR}) harbours the main
implications of the method proposed by Padmanabhan for the
derivation of the field equations in the absence of the derivative of
the Lagrangian density with respect to the metric tensor \cite{Pady}.
To this point, equation (\ref{EquiReLieLE}) could be understood
as the foundation of this method. Apparently, in terms of
Eq. (\ref{LieLEJequiR}), we have demonstrated that each of the
two equations (\ref{LieLagdens}) and (\ref{DivJE}) can be chosen
as the sufficient condition for identifying $\tilde{E}^{\mu\nu}$ with
${E}^{\mu\nu}$, and the spirit for the procedure using the former to
yield ${E}^{\mu\nu}=\tilde{E}^{\mu\nu}$ is in accordance with
that for Padmanabhan's method. Nevertheless, employing the latter
to verify ${E}^{\mu\nu}=\tilde{E}^{\mu\nu}$ possesses the
advantage of getting rid of the derivative of the Lagrangian density
with respect to the metric during the whole process.

On the basis of Eq. (\ref{LieLEJequiR}), let us delve deeper into
the association of Eq. (\ref{ThetaLie}) with both the equations
(\ref{LieLagdens}) and (\ref{DivJE}). According to the first
equivalence relationship in Eq. (\ref{LieLEJequiR}) together with
Eq. (\ref{NSconTheLie2}), the combination of the generalized Bianchi
identity $\nabla_\mu{E}^{\mu\nu}=0$ resulting from Eq. (\ref{LieLagdens})
with Eq. (\ref{DivJE}) enables one to arrive at
\be
\text{Equations }(\ref{LieLagdens})
\text{ and }(\ref{DivJE})\quad\Rightarrow\quad
\Theta^\mu(\delta\rightarrow\mathcal{L}_\zeta)
=2\zeta_\nu{X}^{\mu\nu}
-\nabla_\nu(\text{\textbf{anti-symmetric tensor}})^{\mu\nu}
\, . \label{LieDLtoThet}
\ee
Here the rank-two anti-symmetric tensor contained in the
covariant divergence term obeys
$\nabla_\nu(\text{\textbf{anti-symmetric tensor}})^{\mu\nu}
=\nabla_\nu{Q}^{\mu\nu}$. Apparently, according to
Eq. (\ref{LieDLtoThet}), for the surface term $\Theta^\mu$
arising from the variation of the Lagrangian, the validity for
the two equations (\ref{LieLagdens}) and (\ref{DivJE})
constitutes the sufficient condition for expressing
$\Theta^\mu(\delta\rightarrow\mathcal{L}_\zeta)$
as the form (\ref{ThetaLie}). However, this can be simplified
further. Specifically, note that both the two equations
(\ref{LieLagdens}) and (\ref{DivJE}) originate from the
replacement of the variation operator $\delta$ with the Lie
derivative $\mathcal{L}_\zeta$ within the variation equation
of the Lagrangian (\ref{LagCovR}) just expressed respectively
as two equivalent but formally different equations
(\ref{VarLag}) and (\ref{VarLag3}).
To this point, it is reasonable to conclude that the equivalence
between Eqs. (\ref{VarLag}) and (\ref{VarLag3}) results in the
coexistence of Eqs. (\ref{LieLagdens}) and (\ref{DivJE}). As a
matter of fact, such a conclusion is verified to hold indeed by
means of direct calculations within Appendix \ref{appendB}.
Therefore, from the perspective for the coexistence of
Eqs. (\ref{LieLagdens}) and (\ref{DivJE}), the validity for
either of them can be fully responsible for the decomposition of
the surface term $\Theta^\mu(\delta\rightarrow\mathcal{L}_\zeta)$
into the form (\ref{ThetaLie}). On the other hand, regardless of
the aforementioned derivation for Eq. (\ref{ThetaLie}), in terms
of our process to derive this equation given by Section \ref{three},
it has actually been shown that Eq. (\ref{ThetaLie}) holds
identically within the context of the Lagrangian (\ref{LagCovR}).
Moreover, the existence of this equation can be independent of
Eqs. (\ref{LieLagdens}) and (\ref{DivJE}), attributed to the fact
that Eq. (\ref{ThetaLie}) directly arises from the substitution
of $\delta$ with $\mathcal{L}_\zeta$ in the surface
term $\Theta^\mu$, which merely comes from the sum of the scalar
$Z_{(i)}^{\lambda_1\cdot\cdot\cdot\lambda_i\mu\nu\rho\sigma}
\delta\nabla_{\lambda_1}\cdot\cdot\cdot\nabla_{\lambda_i}
{R}_{\mu\nu\rho\sigma}$ with respect to $i$ from $0$ to $m$
appearing in Eq. (\ref{VariLagdens}).

What is more, due to the analysis mentioned above on the existence
for the form (\ref{ThetaLie}), it is really
unnecessary to particularly stress on that the structure of
$\Theta^\mu(\delta\rightarrow\mathcal{L}_\zeta)$ plays a role in
establishing the equivalence relation between Eqs. (\ref{LieLagdens})
and (\ref{DivJE}) as before. Apparently, as a consequence of
Eqs. (\ref{LieLEJequiR}) and (\ref{LieDLtoThet}),
the setting up of the two equations (\ref{LieLagdens}) and (\ref{DivJE})
automatically brings about the inference that they are completely
equivalent to each other
(\emph{i.e.} \textbf{existence implies equivalence}), namely,
\be
\boxed{\text{Equation }(\ref{LieLagdens})\text{ or }
(\ref{VarLag})|_{\delta\rightarrow\mathcal{L}_\zeta}
\quad\Leftrightarrow\quad
\text{Equation }(\ref{DivJE})\text{ or }
(\ref{VarLag3})|_{\delta\rightarrow\mathcal{L}_\zeta}}
\, . \label{LieLEJequiR2}
\ee
From Eq. (\ref{LieLEJequiR2}), by utilizing Eqs. (\ref{LieLEJequiR})
and (\ref{LieDLtoThet}), we actually have presented a strict proof
that the equivalence relationship between the two forms (\ref{VarLag})
and (\ref{VarLag3}) for the variation of the Lagrangian (\ref{LagCovR})
is still reserved after the variation operator in them is substituted
with the Lie derivative along an arbitrary vector field. Another proof
for Eq. (\ref{LieLEJequiR2}) in terms of straightforward calculations
can be found within Appendix \ref{appendB}. The equivalence between
Eqs. (\ref{LieLagdens}) and (\ref{DivJE}) causes that the equations of
motion arising from Padmanabhan's method have to be in agreement with
our ones from the surface term perspective.

\begin{figure}[h]
\centering
\includegraphics[width=14cm]{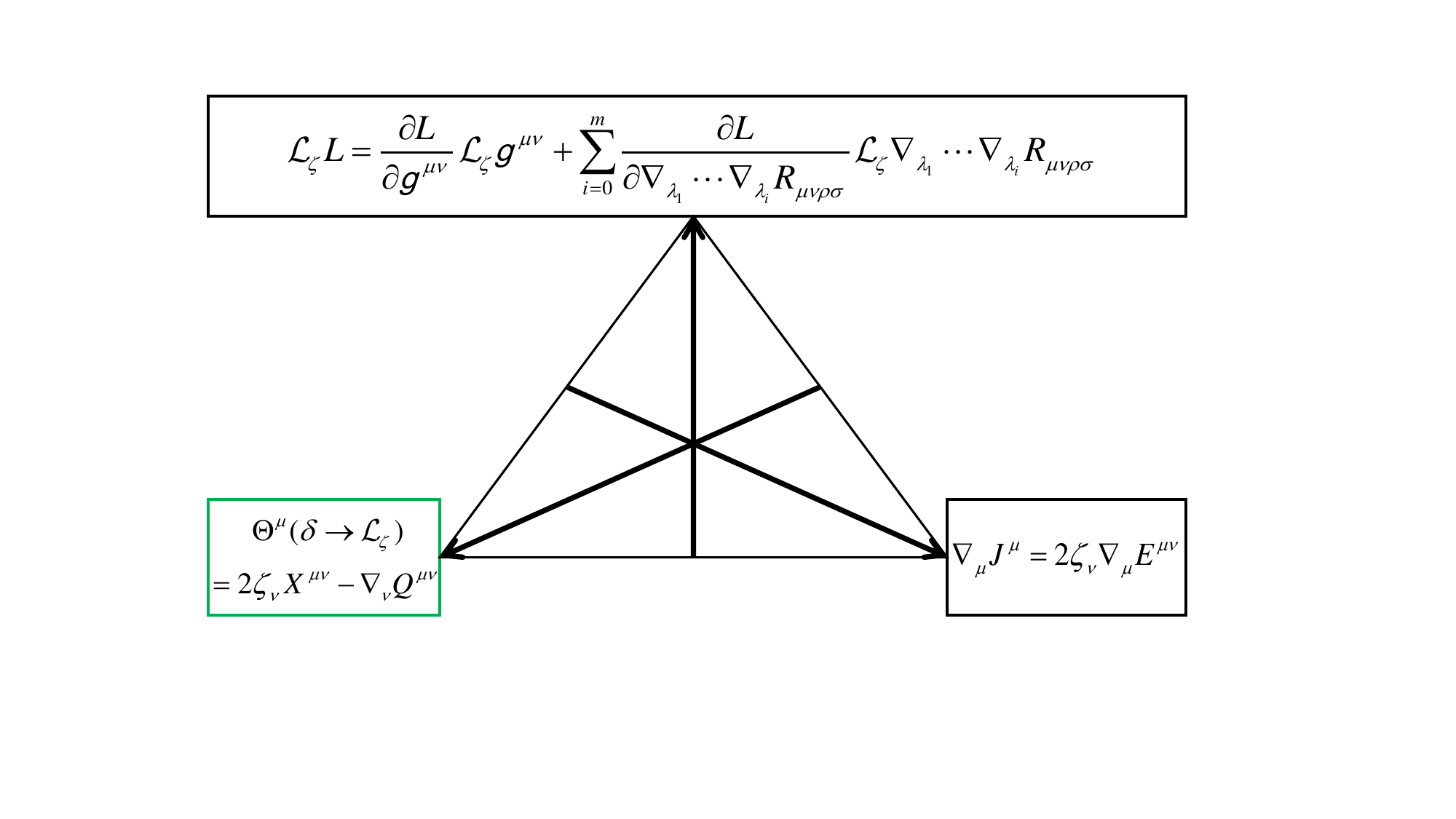}
\vspace{-1.5cm}
\caption{Any one of the three equations
(\ref{ThetaLie}), (\ref{LieLagdens}) and (\ref{DivJE}) is able
to result from the rest two. Equation (\ref{ThetaLie})
holds identically although it can appear as a consequence from the
coexistence of Eqs. (\ref{LieLagdens}) and (\ref{DivJE}). Besides,
the fact that equation (\ref{ThetaLie}) always holds true offers
support for the equivalence between Eqs. (\ref{LieLagdens})
and (\ref{DivJE}).}
\label{FigeEqs}
\end{figure}

Finally, on the basis of the aforementioned analysis in the present
section, we turn to summarizing the relationships among the three
equations (\ref{ThetaLie}), (\ref{LieLagdens}) and (\ref{DivJE}).
They are established under the condition that the variation operator
in Eqs. (\ref{SurfT}), (\ref{VarLag}) and (\ref{VarLag3}) is allowed
to be replaced with the Lie derivative $\mathcal{L}_\zeta$,
respectively. In particular, as has been shown in Appendix
\ref{appendB}, both the two
equations (\ref{LieLagdens}) and (\ref{DivJE}) always coexist
attributed to the equivalence between Eqs. (\ref{VarLag}) and
(\ref{VarLag3}), and they are equivalent under their coexistence.
Besides, the combination of Eqs. (\ref{ThetaLie}) and (\ref{DivJE})
is completely equivalent to the conserved equation for the Noether
current (\ref{ConsCurr}) plus the generalized Bianchi identity for
the Euler-Lagrange expression (\ref{EulerLagEq}), that is,
\be
\boxed{\text{Equations }(\ref{ThetaLie})
\text{ and }(\ref{DivJE})
\quad\Leftrightarrow\quad
\nabla_\mu{E}^{\mu\nu}=0\text{ and }
\nabla_\mu{J}^{\mu}=0}
\, . \label{ThDivJEER}
\ee
Obviously, here Eq. (\ref{ThDivJEER}) provides another way to confirm
the validity of Eqs. (\ref{ThetaLie}) and (\ref{DivJE}). Moreover,
there exists the fact that allowing the substitution
$\delta\rightarrow\mathcal{L}_\zeta$ in any two of Eqs. (\ref{VarLag}),
(\ref{VarLag3}) and (\ref{SurfT}) guarantees that all the three
equations (\ref{ThetaLie}), (\ref{LieLagdens}) and (\ref{DivJE})
must come into existence simultaneously. To see this more clearly,
on the basis of both the two equations (\ref{LieLEJequiR}) and
(\ref{LieDLtoThet}) or according to the two equations
(\ref{RelDivThLieVLag}) and (\ref{EquiRTieLV}), it is easy to find
that the three equations (\ref{ThetaLie}), (\ref{LieLagdens}) and
(\ref{DivJE}) are actually not completely independent to each other
and the arbitrary two equations among them can jointly serve as
the sufficient condition for yielding the remaining one
(see Fig. \ref{FigeEqs} for illustration). In other words, for the
sake of ensuring that all of the three equations are valid, it is
only necessary to confirm that the arbitrary two equations of them hold.

For instance, in order to guarantee that all the three equations
(\ref{ThetaLie}), (\ref{LieLagdens}) and (\ref{DivJE}) hold together,
it is only required to prove that $\nabla_\mu{E}^{\mu\nu}=0$ and
$\nabla_\mu{J}^{\mu}=0$ by using Eq. (\ref{ThDivJEER}). Apart from
this, only from the perspective of the coexistence relationship between
the two equations (\ref{LieLagdens}) and (\ref{DivJE}), a simpler way
is to verify that Eq. (\ref{LieLagdens}) or (\ref{DivJE}) is valid
since the validity for either of them indicates that the other one
is valid as well. In view of
Eq. (\ref{EquiReLieLE}), one feasible way to accomplish this
is to test directly that both the two equations (\ref{CovdeL})
and (\ref{Partgmnequal}) hold simultaneously. What is more,
it has been demonstrated that the surface term $\Theta^\mu$ with
$\delta\rightarrow\mathcal{L}_\zeta$ can be put into the form
(\ref{ThetaLie}) via the straightforward calculations having
nothing to do with Eqs. (\ref{LieLagdens}) and (\ref{DivJE}).
This implies that Eq. (\ref{ThetaLie}) can exist independently
of these two equations. However, the coexistence of
Eqs. (\ref{LieLagdens}) and (\ref{DivJE}) is able to produce
Eq. (\ref{ThetaLie}). So the existence of Eq. (\ref{ThetaLie})
is a necessary condition for guaranteeing that both the equations
(\ref{LieLagdens}) and (\ref{DivJE}) hold simultaneously. On the
other hand, although the existence for Eq. (\ref{ThetaLie}) is not
a sufficient condition for determining that either of
Eqs. (\ref{LieLagdens}) and (\ref{DivJE}) definitely
holds, equation (\ref{ThetaLie}) can be employed to assist us to
derive either of them out of the other one.

\section{The disappearance for the covariant divergence of
the expression of the field equations}\label{six}

As is by now well-known, within the framework of general relativity,
in order to ensure the conservation law of energy-momentum tensor
to hold, the vanishing for the covariant divergence of Einstein
tensor is viewed as one of the fundamental assumptions in curved
spacetimes. In the same way, the problem for conservation law
is of great importance as well within the context of a wide
variety of theories for gravity admitting diffeomorphism invariance
beyond general relativity. Owing to this, it is
necessary for us to take into account the covariant divergence of
the expression of the field equations relative to the Lagrangian
(\ref{LagCovR}). However, unlike the situation for general relativity,
the related calculations become more complicated as a consequence for
the appearance of the covariant derivatives of the Rienmann
curvature tensor.

Due to the above, it can be found that there exist at least four
ways towards the proof for the generalized Bianchi identity
$\nabla_\mu{E}^{\mu\nu}=0$. (\textbf{I}) According to
Eq. (\ref{ProceOne}), the first way is to adopt the identity
(\ref{Partgmnequal}) as the starting point to yield the off-shell
Noether current ${J}^\mu=\nabla_\nu{Q}^{\mu\nu}$, which further
gives rise to $\nabla_\mu{J}^\mu=0$. Then Eq. (\ref{DivJE}) leads
to $\nabla_\mu{E}^{\mu\nu}=0$. (\textbf{II}) In light of
Eq. (\ref{EJequivR}), the second way is to directly verify the
vanishing of the covariant divergence of ${J}^\mu$.
(\textbf{III}) Under the condition that Eq. (\ref{DivJE}) holds true,
the third way is to utilize the conclusion that the surface term
$\Theta^\mu(\delta\rightarrow\mathcal{L}_\zeta)$ can be expressed
as the form (\ref{ThetLieGF}), directly yielding
$\nabla_\mu{E}^{\mu\nu}=0$ by means of Eq. (\ref{NSconTheLie}).
(\textbf{IV}) The fourth way is to
employ Eq. (\ref{LieLagdens}) as the starting point. Such an
equation results in two identities (\ref{CovdeL}) and
(\ref{Partgmnequal}). According to Eq. (\ref{ProceOne}), the latter
enables one to obtain ${E}^{\mu\nu}=\tilde{E}^{\mu\nu}$. In terms
of this, we directly compute the covariant divergence of
$\tilde{E}^{\mu\nu}$ to produce $\nabla_\mu\tilde{E}^{\mu\nu}=0$
by the aid of the identity (\ref{CovdeL}). Actually, all the four
ways are not completely independent to each other, and either of
Eqs. (\ref{LieLagdens}) and (\ref{DivJE}) is able to serve as the
starting point.

In spite of the fact that the generalized Bianchi identity can be
successfully proved via each of the first three ways, in the remainder
of this section, we will follow the fourth way to perform straightforward
calculations on the covariant divergence of the expression for equations
of motion to confirm that it indeed vanishes identically within the context
of matter free gravity theories endowed with the Lagrangian (\ref{LagCovR}).
All the related calculations are based on the expression (\ref{EoMtild})
for equations of motion arising from the surface term rather than the
one (\ref{EulerLagEq}) corresponding to the Euler-Lagrange equations.

As a beginning, we handle the covariant divergence of the
${X}^{\mu\nu}$ tensor. The straightforward calculations bring forth
\bea
\nabla_\mu{X}^{\mu\nu}&=&
R^{\nu}_{~\lambda\rho\sigma}
\nabla_\mu{P}^{\mu\lambda\rho\sigma}
+{P}^{\mu\lambda\rho\sigma}
\nabla_\mu R^{\nu}_{~\lambda\rho\sigma}
-2\nabla_\mu\nabla_{\rho}\nabla_{\sigma}
{P}^{\rho\mu\nu\sigma}
+\nabla_\mu{W}^{\mu\nu}
\, . \label{DivXmn}
\eea
Within the above equation, the vector ${P}^{\mu\lambda\rho\sigma}
\nabla_\mu R^{\nu}_{~\lambda\rho\sigma}$ is alternatively expressed
as
\be
{P}^{\mu\lambda\rho\sigma}
\nabla_\mu R^{\nu}_{~\lambda\rho\sigma}
=\frac{1}{2}{P}^{\alpha\beta\rho\sigma}
\nabla^\nu R_{\alpha\beta\rho\sigma}
\, , \label{PDivRiem}
\ee
and the vector $\nabla_{\mu}\nabla_{\rho}
\nabla_\sigma{P}^{\rho\mu\nu\sigma}$ can be rewritten as
\be
\nabla_{\mu}\nabla_{\rho}\nabla_\sigma
{P}^{\rho\mu\nu\sigma}=-\nabla_{[\rho}\nabla_{\mu]}
\nabla_\sigma{P}^{\rho\mu\nu\sigma}=
\frac{1}{2}R^{\nu}_{~\lambda\rho\sigma}
\nabla_\mu{P}^{\mu\lambda\rho\sigma}
\, . \label{Divdel2Pijp}
\ee
In order to figure out the divergence of the tensor
${W}^{\mu\nu}=\sum^m_{i=0}{W}^{\mu\nu}_{(i)}$, a primary task could
be to deal with $\nabla_\mu{W}^{\mu\nu}_{(i)}$, yielding
\bea
\nabla_\mu{W}^{\mu\nu}_{(i)}&=&
\frac{1}{2}\sum_{k=1}^i(-1)^{k-1}
\left(\nabla_{\lambda_{k-1}}
\cdot\cdot\cdot\nabla_{\lambda_{1}}
Z_{(i)}^{\lambda_1\cdot\cdot\cdot\lambda_{i}
\alpha\beta\rho\sigma}\right)\nabla^\nu
\nabla_{\lambda_k}
\cdot\cdot\cdot\nabla_{\lambda_{i}}
R_{\alpha\beta\rho\sigma} \nn \\
&&-\frac{1}{2}\sum_{k=1}^i(-1)^{k}
\left(\nabla_{\lambda_k}
\cdot\cdot\cdot\nabla_{\lambda_{1}}
Z_{(i)}^{\lambda_1\cdot\cdot\cdot\lambda_{i}
\alpha\beta\rho\sigma}\right)\nabla^\nu
\nabla_{\lambda_{k+1}}
\cdot\cdot\cdot\nabla_{\lambda_{i}}
R_{\alpha\beta\rho\sigma}  \nn \\
&&-\frac{1}{2}R^\nu_{~\lambda\rho\sigma}
{U}^{\rho\lambda\sigma}_{(i)}
+\frac{1}{2}\nabla_\mu\nabla_\lambda
\left({U}^{(\mu\nu)\lambda}_{(i)}
-{U}^{\lambda(\mu\nu)}_{(i)}
+{U}^{[\mu|\lambda|\nu]}_{(i)}\right)
\, . \label{DivWimn}
\eea
We observe that the vector $R^\nu_{~\lambda\rho\sigma}
{U}^{\rho\lambda\sigma}_{(i)}$ in Eq. (\ref{DivWimn}) is able to
be further expressed as
\bea
R^\nu_{~\lambda\rho\sigma}
{U}^{\rho\lambda\sigma}_{(i)}&=&
\frac{1}{2}\big(3R^\nu_{~[\sigma\rho\lambda]}
+2R^\nu_{~\lambda\rho\sigma}\big)
{U}^{\rho\lambda\sigma}_{(i)} \nn \\
&=&\frac{1}{2}\big(R^\nu_{~\sigma\rho\lambda}
+R^\nu_{~\lambda\rho\sigma}
+R^\nu_{~\rho\lambda\sigma}\big)
{U}^{\rho\lambda\sigma}_{(i)} \nn \\
&=&\nabla_\mu\nabla_\lambda
\left({U}^{(\mu\nu)\lambda}_{(i)}
-{U}^{\lambda(\mu\nu)}_{(i)}
+{U}^{[\mu|\lambda|\nu]}_{(i)}\right)
\, . \label{DivUi}
\eea
Then substituting Eq. (\ref{DivUi}) into (\ref{DivWimn}) leads to
\be
\nabla_\mu{W}^{\mu\nu}_{(i)}=
\frac{1}{2}{Z}_{(i)}^{\lambda_1\cdot\cdot\cdot\lambda_{i}
\alpha\beta\rho\sigma}\nabla^\nu
\nabla_{\lambda_1}
\cdot\cdot\cdot\nabla_{\lambda_{i}}
R_{\alpha\beta\rho\sigma}
-\frac{1}{2}(-1)^{i}\Big(\nabla_{\lambda_i}
\cdot\cdot\cdot\nabla_{\lambda_{1}}
Z_{(i)}^{\lambda_1\cdot\cdot\cdot\lambda_{i}
\alpha\beta\rho\sigma}\Big)
\nabla^\nu R_{\alpha\beta\rho\sigma}
\, . \label{DivWimn2}
\ee
As a result of the sum for equation (\ref{DivWimn2}) with respect to
$i$ from 0 to $m$, the covariant divergence of the ${W}^{\mu\nu}$
tensor is read off as
\be
\nabla_\mu{W}^{\mu\nu}=
\frac{1}{2}\sum^m_{i=0}{Z}^{\lambda_1\cdot\cdot\cdot\lambda_{i}
\alpha\beta\rho\sigma}_{(i)}\nabla^\nu
\nabla_{\lambda_1}
\cdot\cdot\cdot\nabla_{\lambda_{i}}
R_{\alpha\beta\rho\sigma}
-\frac{1}{2}P^{\alpha\beta\rho\sigma}
\nabla^\nu{R}_{\alpha\beta\rho\sigma}
\, . \label{DivWmn}
\ee
In terms of Eqs. (\ref{PDivRiem}), (\ref{Divdel2Pijp}) and (\ref{DivWmn}),
plugging them back into equation (\ref{DivXmn}) gives rise to
\bea
\nabla_\mu{X}^{\mu\nu}&=&
\frac{1}{2}\sum^m_{i=0}{Z}^{\lambda_1\cdot\cdot\cdot\lambda_{i}
\alpha\beta\rho\sigma}_{(i)}\nabla^\nu
\nabla_{\lambda_1}
\cdot\cdot\cdot\nabla_{\lambda_{i}}
R_{\alpha\beta\rho\sigma}
\, . \label{DivXmn2}
\eea
Here we point out that both the two equations (\ref{DivWmn}) and
(\ref{DivXmn2}) hold identically and they depend upon the
${Z}^{\lambda_1\cdot\cdot\cdot\lambda_{i} \mu\nu\rho\sigma}_{(i)}$
tensors.

With Eq. (\ref{DivXmn2}) at hand, combining it with equation
(\ref{CovdeL}) assists us to arrive ultimately at the
Bianchi-type identity
\be
\nabla_\mu{E}^{\mu\nu}=\nabla_\mu\tilde{E}^{\mu\nu}=
\nabla_\mu{X}^{\mu\nu}-\frac{1}{2}\nabla^\nu{L}
=0 \, . \label{DivEmn}
\ee
Consequently, we prove that ${E}^{\mu\nu}$ is indeed divergence-free.

Finally, we point out that the vector $\nabla_\mu{W}^{\mu\nu}$ can
be derived out of Eq. (\ref{ZidelnabR}) under the replacement
$\delta\rightarrow\mathcal{L}_\zeta$, that is,
\bea
Z_{(i)}^{\lambda_1\cdot\cdot\cdot\lambda_i\mu\nu\rho\sigma}
\mathcal{L}_\zeta\nabla_{\lambda_1}
\cdot\cdot\cdot\nabla_{\lambda_i}{R}_{\mu\nu\rho\sigma}&=&
(-1)^i\big(\nabla_{\lambda_{i}}\cdot\cdot\cdot\nabla_{\lambda_1}
Z_{(i)}^{\lambda_1\cdot\cdot\cdot\lambda_i\mu\nu\rho\sigma}\big)
\mathcal{L}_\zeta{R}_{\mu\nu\rho\sigma} \nn \\
&&-2{B}^{\mu\nu}_{(i)}\nabla_{\mu}\zeta_{\nu}
+\nabla_\mu\bar{\Theta}^\mu_{(i)}(\delta\rightarrow\mathcal{L}_\zeta)
\, . \label{ZidelnabRLie}
\eea
Specifically, substituting Eq. (\ref{BarThetLie}) into
(\ref{ZidelnabRLie}) and summing the result with respect to $i$,
we have
\bea
2\zeta_\nu\nabla_\mu{W}^{\mu\nu}&=&
2\big({B}^{\mu\nu}-{W}^{\mu\nu}\big)\nabla_\mu\zeta_\nu
-P^{\alpha\beta\rho\sigma}\mathcal{L}_\zeta{R}_{\alpha\beta\rho\sigma}
\nn \\
&&+\sum^{m}_{i=0}
{Z}_{(i)}^{\lambda_1\cdot\cdot\cdot\lambda_i\mu\nu\rho\sigma}
\mathcal{L}_\zeta\nabla_{\lambda_1}
\cdot\cdot\cdot\nabla_{\lambda_i}{R}_{\mu\nu\rho\sigma}
\, . \label{ZPDivWLie}
\eea
The above equation can be recast into
\be
\zeta_\nu\left(2\nabla_\mu{W}^{\mu\nu}+{P}^{\alpha\beta\rho\sigma}
\nabla^\nu{R}_{\alpha\beta\rho\sigma}
-\sum^m_{i=0}{Z}^{\lambda_1\cdot\cdot\cdot\lambda_{i}
\alpha\beta\rho\sigma}_{(i)}\nabla^\nu
\nabla_{\lambda_1}
\cdot\cdot\cdot\nabla_{\lambda_{i}}
R_{\alpha\beta\rho\sigma}\right)
+\Upsilon^{\mu\nu}\nabla_\mu\zeta_\nu=0
\, , \label{ZPDivWLie2}
\ee
where the rank-two tensor $\Upsilon^{\mu\nu}$ is defined through
 \bea
\Upsilon^{\mu\nu}&=&
-\sum^m_{i=0}\sum^i_{k=1}
{Z}^{\lambda_1\cdot\cdot\cdot\lambda_{k-1}\mu
\lambda_{k+1}\cdot\cdot\cdot\lambda_{i}\alpha\beta\rho\sigma}_{(i)}
\nabla_{\lambda_{1}}\cdot\cdot\cdot
\nabla_{\lambda_{k-1}}\nabla^\nu \nabla_{\lambda_{k+1}}
\cdot\cdot\cdot\nabla_{\lambda_{i}}R_{\alpha\beta\rho\sigma} \nn \\
&&-4\sum^{m}_{i=0}
{Z}^{\lambda_1\cdot\cdot\cdot\lambda_{i}\mu\lambda\rho\sigma}_{(i)}
\nabla_{\lambda_{1}}\cdot\cdot\cdot\nabla_{\lambda_{i}}
{R}^{\nu}_{~\lambda\rho\sigma}+4{P}^{\mu\lambda\rho\sigma}
{R}^{\nu}_{~\lambda\rho\sigma}+2{W}^{\mu\nu}-2{B}^{\mu\nu}
\, . \label{Upsimndef}
\eea
Within Eq. (\ref{ZPDivWLie2}), since this equation is valid for any
vector field $\zeta^\mu$, the vector field proportional to $\zeta^\mu$
and the tensor $\Upsilon^{\mu\nu}$ have to vanish together. The former
conclusion gives rise to Eq. (\ref{DivWmn}) for $\nabla_\mu{W}^{\mu\nu}$,
while $\Upsilon^{\mu\nu}=0$ reproduces the equality (\ref{PWBexpan}).
On the other hand, by means of the combination of Eq. (\ref{LieLagdens})
with (\ref{ZPDivWLie}), we obtain
\bea
&&\zeta_\nu\left(2\nabla_\mu{W}^{\mu\nu}+{P}^{\alpha\beta\rho\sigma}
\nabla^\nu{R}_{\alpha\beta\rho\sigma}
-\nabla^\nu{L}\right) \nn \\
&&+2\left(2{P}^{\mu\lambda\rho\sigma}
{R}^{\nu}_{~\lambda\rho\sigma}+{W}^{\mu\nu}-{B}^{\mu\nu}
-\frac{\partial{L}}{\partial{g}^{\rho\sigma}}
{g}^{\mu\rho}{g}^{\nu\sigma}\right)\nabla_\mu\zeta_\nu=0
\, . \label{ZPDivWLie3}
\eea
In view of the arbitrariness of the vector field $\zeta^\mu$
in Eq. (\ref{ZPDivWLie3}), another form for $\nabla_\mu{W}^{\mu\nu}$,
resulting from the vanishing of the vector contracted with $\zeta_\nu$,
is read off as
\be
\nabla_\mu{W}^{\mu\nu}=
\frac{1}{2}\nabla^\nu{L}-\frac{1}{2}{P}^{\alpha\beta\rho\sigma}
\nabla^\nu{R}_{\alpha\beta\rho\sigma}
\, . \label{DivWmn2}
\ee
Meanwhile, apart from Eq. (\ref{DivWmn2}), the disappearance of the
rank-two tensor contracted with $\nabla_\mu\zeta_\nu$ leads to the
equality (\ref{Partgmnequal}) once again.

\section{The procedure for extracting the expression of the field
equations out of the surface term}\label{seven}

According to the aforementioned analysis, equation (\ref{LieLEJequiR})
enables us to conclude that \emph{the validity for either of
Eqs. (\ref{LieLagdens}) and (\ref{DivJE})
is a sufficient and necessary condition for that (\textbf{I}) the
${X}^{\mu\nu}$ tensor arising from the surface term can be employed
to define the expression ${E}^{\mu\nu}$ for the equations of motion
as ${E}^{\mu\nu}={X}^{\mu\nu}-\frac{1}{2}{L}{g}^{\mu\nu}$ and
(\textbf{II}) the covariant divergence of the resulting ${E}^{\mu\nu}$
satisfies $\nabla_\mu{E}^{\mu\nu}=0$.}

In terms of the conclusion mentioned above together with
Eq. (\ref{FourEquiv}), the process (\ref{ProcST}) devoted to the
derivation for the expression of the field equations from the surface
term perspective is concretized into the following. \\
\boxed{\textbf{Step 1}}: Varying the Lagrangian density ${L}$ instead
of the Lagrangian $\mathcal{L}$ with respect to all the variables
gives rise to Eq. (\ref{VariLagdens}). Note that such an equation
describing the variation for the Lagrangian density constitutes the
starting point of the entire process. Subsequently, on the basis of
Eq. (\ref{VariLagdens}), regardless of any term associated with the
variation of the metric tensor, handling all the terms involving the
variation of the variables
$\nabla_{\lambda_1}\cdot\cdot\cdot\nabla_{\lambda_i}
{R}_{\mu\nu\rho\sigma}$s $(i=0,\cdot\cdot\cdot,m)$ leads to the
surface term $\Theta^\mu$.\\
\boxed{\textbf{Step 2}}: In terms of the surface term $\Theta^\mu$, under
the transformation that the variation operator $\delta$ is replaced
with the Lie derivative $\mathcal{L}_\zeta$ along an arbitrary vector
field $\zeta^\mu$, this surface term is able to be decomposed into
the form (\ref{ThetaLie}), which generates the rank-two tensor
${X}^{\mu\nu}$ together with the anti-symmetric one ${Q}^{\mu\nu}$.
As a matter of fact, decomposing the surface term
with $\delta\rightarrow\mathcal{L}_\zeta$ into the form given by
Eq. (\ref{ThetLieGF}) can be regarded as the crucial step towards
extracting the expression for the equations of motion.\\
\boxed{\textbf{Step 3}}: In terms of the expression (\ref{ThetaLie}) for
the surface term $\Theta^\mu$ with $\delta\rightarrow\mathcal{L}_\zeta$,
under the condition that equation (\ref{VariLagdens}) with $\delta$
substituted by $\mathcal{L}_\zeta$, presented by
Eq. (\ref{LieLagdens}), is supposed to be valid, or under the condition
that equation (\ref{DivJE}) holds, the expression ${E}^{\mu\nu}$ for
the field equations derived out of the variation of the Lagrangian
can be built from the ${X}^{\mu\nu}$ tensor, written as
${E}^{\mu\nu}={X}^{\mu\nu}-\frac{1}{2}{L}{g}^{\mu\nu}$. Here the validity
for either of Eqs. (\ref{LieLagdens}) and (\ref{DivJE}) supplies a
guarantee of the symmetry for the rank-two tensor ${X}^{\mu\nu}$, namely,
${X}^{\mu\nu}={X}^{\nu\mu}$. Meanwhile, the combination of ${E}^{\mu\nu}$
with the quantity $\Theta^\mu(\delta\rightarrow\mathcal{L}_\zeta)$
leads to the Noether current ${J}^\mu$, taking the form
${J}^\mu=2\zeta_\nu{E}^{\mu\nu}+\zeta^\mu{L}
-\Theta^\mu(\delta\rightarrow\mathcal{L}_\zeta)$
or ${J}^\mu=\nabla_\nu{Q}^{\mu\nu}$.

Moreover, inspired with the three steps mentioned above, we find
that the variation of the Lagrangian density together with its Lie
derivative along an arbitrary vector field can be sufficient for
producing the expressions of the field equations and the Noether
current in a uniform framework. In terms of this, as an application
to fully illustrate the surface term perspective towards the
derivation for the expression of the field equations proposed above,
we demonstrate within the remainder of this section that all the results
related to the Lagrangian (\ref{LagCovR}) can be directly generalized
to the following Lagrangian 
\be
\sqrt{-g}{L}_{gen}=\sqrt{-g}{L}_{gen}\left({g}^{\mu\nu},
{R}_{\alpha\beta\rho\sigma},\nabla_\lambda\right)
\, . \label{LagGen}
\ee
Here the Lagrangian (\ref{LagGen}) is more general than the one
given by Eq. (\ref{LagCovR}) in form. However, each of them can
be employed to describe the theories of gravity depending on the
metric and the covariant derivatives of the Riemann tensor.

Specifically, for such gravity theories armed with the Lagrangian
(\ref{LagGen}), as a starting point, we only need to compute the
variation of the Lagrangian density ${L}_{gen}$. In general, the
result can be always put into the following form
\be
\delta{L}_{gen}=\check{S}_{\mu\nu}\delta{g}^{\mu\nu}
+\sum^{m}_{i=0}
\check{Z}_{(i)}^{\lambda_1\cdot\cdot\cdot\lambda_i\mu\nu\rho\sigma}
\delta\nabla_{\lambda_1}
\cdot\cdot\cdot\nabla_{\lambda_i}{R}_{\mu\nu\rho\sigma}
\, . \label{VaricheckL}
\ee
Within Eq. (\ref{VaricheckL}), the nonnegative integer $m$ stands for
the maximum number of the covariant derivative operator acting on
the Riemann curvature tensor, while the symmetric rank-two tensor
$\check{S}_{\mu\nu}$ and the rank-$(i+4)$ tensors
$\check{Z}_{(i)}^{\lambda_1\cdot\cdot\cdot\lambda_i\mu\nu\rho\sigma}$s
are actually allowed to be arbitrary in form although they are
constrained by the variation of the Lagrangian density. For convenience,
like before, we impose the restriction that the last four indices in the
$\check{Z}_{(i)}^{\lambda_1\cdot\cdot\cdot\lambda_i\mu\nu\rho\sigma}$
tensor identically satisfy
$\check{Z}_{(i)}^{\lambda_1\cdot\cdot\cdot\lambda_i\mu\nu\rho\sigma}=
\Delta^{\mu\nu\rho\sigma}_{\alpha\beta\gamma\lambda}
\check{Z}_{(i)}^{\lambda_1\cdot\cdot\cdot\lambda_i\alpha\beta\gamma\lambda}$
attributed to the relation $\delta\nabla_{\lambda_1}
\cdot\cdot\cdot\nabla_{\lambda_i}{R}_{\mu\nu\rho\sigma}=
\Delta^{\alpha\beta\gamma\lambda}_{\mu\nu\rho\sigma}
\delta\nabla_{\lambda_1}
\cdot\cdot\cdot\nabla_{\lambda_i}{R}_{\alpha\beta\gamma\lambda}$.
Thanks to the fact that
the surface term $\Theta^\mu_{gen}$ is adequately determined by the
second term at the right hand side of Eq. (\ref{VaricheckL}), it is
only necessary to deal with this term regardless of the scalar
$\check{S}_{\mu\nu}\delta{g}^{\mu\nu}$ for the sake of deriving the
expression for the equations of motion out of the surface term
$\Theta^\mu_{gen}$. As a consequence of doing so, the scalar
$\sum^{m}_{i=0}
\check{Z}_{(i)}^{\lambda_1\cdot\cdot\cdot\lambda_i\mu\nu\rho\sigma}
\delta\nabla_{\lambda_1}
\cdot\cdot\cdot\nabla_{\lambda_i}R_{\mu\nu\rho\sigma}$
is connected with the divergence of
the surface term $\Theta^\mu_{gen}$ in the way
\be
\sum^{m}_{i=0}
\check{Z}_{(i)}^{\lambda_1\cdot\cdot\cdot\lambda_i\mu\nu\rho\sigma}
\delta\nabla_{\lambda_1}
\cdot\cdot\cdot\nabla_{\lambda_i}R_{\mu\nu\rho\sigma}
=\check{N}_{\mu\nu}\delta{g}^{\mu\nu}
+\nabla_\mu\Theta^\mu_{gen}
\, , \label{DivThetLgen}
\ee
where the rank-two symmetric tensor $\check{N}_{\mu\nu}$ is given by
\be
\check{N}^{\mu\nu}=\check{B}^{\mu\nu}
-\check{P}^{\mu\lambda\rho\sigma}
{R}^{\nu}_{~\lambda\rho\sigma}
-2\nabla_\rho\nabla_\sigma\check{P}^{\rho\mu\nu\sigma}
\, , \label{checNdef}
\ee
with $\check{B}^{\mu\nu}={B}^{\mu\nu}|_{{Z}\rightarrow\check{Z}}$
and $\check{P}^{\mu\nu\rho\sigma}={P}^{\mu\nu\rho\sigma}
|_{{Z}\rightarrow\check{Z}}$, while the surface term
$\Theta^\mu_{gen}=\Theta^\mu|_{{Z}\rightarrow\check{Z}}$
takes the explicit form
\be
\Theta^\mu_{gen}=2\check{P}^{\mu\nu\rho\sigma}
\nabla_\sigma\delta{g}_{\rho\nu}
-2(\delta{g}_{\nu\rho})\nabla_\sigma\check{P}^{\mu\nu\rho\sigma}
+\sum^{m}_{i=0}\bar{\Theta}^\mu_{(i)}\big|_{{Z}\rightarrow\check{Z}}
\, , \label{ThetLgen}
\ee
with $\bar{\Theta}^\mu_{(i)}$ given by Eq. (\ref{Thetbar}).
Here it is important to emphasize that equation (\ref{DivThetLgen})
holds true for any nonnegative integer $m$ and arbitrary tensors
$\check{Z}_{(i)}^{\lambda_1\cdot\cdot\cdot\lambda_i\mu\nu\rho\sigma}$s
regardless of whether such tensors arise as the results from the
variation of the Lagrangian density ${L}_{gen}$ or not.

Moreover, by analogy with the derivation for the identity
(\ref{SumZiLieR}), replacing the variation operator $\delta$ in
the scalar $\sum^{m}_{i=0}
\check{Z}_{(i)}^{\lambda_1\cdot\cdot\cdot\lambda_i\mu\nu\rho\sigma}
\delta\nabla_{\lambda_1}
\cdot\cdot\cdot\nabla_{\lambda_i}R_{\mu\nu\rho\sigma}$ with the Lie
derivative $\mathcal{L}_\zeta$ along the arbitrary smooth vector field
$\zeta^\mu$ results in the decomposition that this scalar is written
as the linear combination of a term proportional to
$\mathcal{L}_\zeta{g}^{\mu\nu}$ with a divergence term, namely,
\be
\sum^{m}_{i=0}
\check{Z}_{(i)}^{\lambda_1\cdot\cdot\cdot\lambda_i\mu\nu\rho\sigma}
\mathcal{L}_\zeta\nabla_{\lambda_1}
\cdot\cdot\cdot\nabla_{\lambda_i}{R}_{\mu\nu\rho\sigma}
=\check{N}_{\mu\nu}\mathcal{L}_\zeta{g}^{\mu\nu}
+2\nabla_\mu\left(\zeta_\nu\check{X}^{\mu\nu}\right)
\, . \label{SumCheZiLieR}
\ee
Within the above equation, the rank-two tensor $\check{X}^{\mu\nu}$
is defined as
\be
\check{X}^{\mu\nu}={X}^{\mu\nu}|_{{Z}\rightarrow\check{Z}}
=\check{P}^{\mu\lambda\rho\sigma}
{R}^{\nu}_{~\lambda\rho\sigma}-2\nabla_{\rho}\nabla_{\sigma}
\check{P}^{\rho\mu\nu\sigma}+\check{W}^{\mu\nu}
\, . \label{checXdef}
\ee
Here $\check{W}^{\mu\nu}=\sum^{m}_{i=0}
{W}^{\mu\nu}_{(i)}\big|_{{Z}\rightarrow\check{Z}}$
with the $W^{\mu\nu}_{(i)}$ tensor given by Eq. (\ref{Widef}).
After a straightforward calculation, the covariant divergence of
the $\check{X}^{\mu\nu}$ tensor turns into
\be
\nabla_\mu\check{X}^{\mu\nu}=\frac{1}{2}\sum^{m}_{i=0}
\check{Z}_{(i)}^{\lambda_1\cdot\cdot\cdot\lambda_i\alpha\beta\rho\sigma}
\nabla^\nu\nabla_{\lambda_1}
\cdot\cdot\cdot\nabla_{\lambda_i}{R}_{\alpha\beta\rho\sigma}
\, . \label{CdivchecX}
\ee
Equation (\ref{CdivchecX}) is a necessary condition for the validity
of the identity (\ref{SumCheZiLieR}).
In the same way, the identity (\ref{SumCheZiLieR}) holds for any
$\check{Z}_{(i)}^{\lambda_1\cdot\cdot\cdot\lambda_i\mu\nu\rho\sigma}$
tensor.

What is more, we observe that the pair
$\big(\check{N}^{\mu\nu},\check{X}^{\mu\nu}\big)$ appearing at the
right hand side of the identity (\ref{SumCheZiLieR}) is non-unique.
Here both the rank-two tensors $\check{N}^{\mu\nu}$ and
$\check{X}^{\mu\nu}$ denote the symmetric tensor proportional to
$\mathcal{L}_\zeta{g}^{\mu\nu}$ and the one contracted with
$\zeta_\nu$ in the divergence term, respectively. For the sake of seeing
this clearly, in addition to the pair
$\big(\check{N}^{\mu\nu},\check{X}^{\mu\nu}\big)$, it is assumed that
there exists another arbitrary pair for rank-two tensors
$\big(\check{\mathcal{N}}^{\mu\nu}, \check{\mathcal{X}}^{\mu\nu}\big)$
being independent of the vector field
$\zeta^\mu$, in which the $\check{\mathcal{N}}^{\mu\nu}$ tensor is
required additionally to be symmetric. It can be proved that a
\textbf{sufficient and necessary condition} for
\bea
\sum^{m}_{i=0}
\check{Z}_{(i)}^{\lambda_1\cdot\cdot\cdot\lambda_i\mu\nu\rho\sigma}
\mathcal{L}_\zeta\nabla_{\lambda_1}
\cdot\cdot\cdot\nabla_{\lambda_i}{R}_{\mu\nu\rho\sigma}
&=&\check{N}_{\mu\nu}\mathcal{L}_\zeta{g}^{\mu\nu}
+2\nabla_\mu\left(\zeta_\nu\check{X}^{\mu\nu}\right) \nn \\
&=&\check{\mathcal{N}}_{\mu\nu}\mathcal{L}_\zeta{g}^{\mu\nu}
+2\nabla_\mu\left(\zeta_\nu\check{\mathcal{X}}^{\mu\nu}\right)
\,  \label{SumCheZiLieR2}
\eea
is that there exists a certain rank-two symmetric and divergence-free
tensor $\Omega^{\mu\nu}$ that is irrelevant to the vector field
$\zeta^\mu$, namely, this tensor has to fulfill
\be
\Omega^{\mu\nu}=\Omega^{\nu\mu}\, , \qquad
\nabla_\mu\Omega^{\mu\nu}=0
\, , \label{PropOmeg}
\ee
rendering the pair $\big(\check{N}^{\mu\nu},\check{X}^{\mu\nu}\big)$
connected with the one
$\big(\check{\mathcal{N}}^{\mu\nu}, \check{\mathcal{X}}^{\mu\nu}\big)$
in the following manner
\be
\check{N}^{\mu\nu}=\check{\mathcal{N}}^{\mu\nu}+\Omega^{\mu\nu}
\, , \qquad
\check{X}^{\mu\nu}=\check{\mathcal{X}}^{\mu\nu}+\Omega^{\mu\nu}
\, . \label{cheNXtrans}
\ee
As a matter of fact, it is easy to see that there are infinite number
of rank-two tensors satisfying the condition (\ref{PropOmeg}), which
accordingly bring forth an infinite number of pairs
$\big(\check{\mathcal{N}}^{\mu\nu}, \check{\mathcal{X}}^{\mu\nu}\big)$
fulfilling Eq. (\ref{SumCheZiLieR}). However, according to
Eq. (\ref{cheNXtrans}), once either of $\check{\mathcal{N}}^{\mu\nu}$
and $\check{\mathcal{X}}^{\mu\nu}$ is fixed, the other one of them
must be determined uniquely. This implies that
$\check{\mathcal{N}}^{\mu\nu}$ and $\check{\mathcal{X}}^{\mu\nu}$ are
in one-to-one correspondence. Therefore, within Eq. (\ref{SumCheZiLieR}),
when the symmetric tensor $\check{N}^{\mu\nu}$ is constrained to inherit
the one proportional to the variation of the metric in
Eq. (\ref{DivThetLgen}), which is given by Eq. (\ref{checNdef}),
the $\check{X}^{\mu\nu}$ tensor takes the unique form (\ref{checXdef}).
Apparently, owing to the fact that the $\Omega^{\mu\nu}$ tensor is
symmetric and divergence-free, it has the potential to convert
to the expression for the equations of motion associated to the 
theories of pure gravity endowed with diffeomorphism invariance 
if another appropriate pair is supplied in addition to the one 
$\big(\check{N}^{\mu\nu},\check{X}^{\mu\nu}\big)$. This will be 
seen below.

In terms of the surface term $\Theta^\mu_{gen}$, the straightforward
computation demonstrates that the counterpart of this term under the
substitution $\delta\rightarrow\mathcal{L}_\zeta$ accordingly exhibits
the form
\be
\boxed{\Theta^\mu_{gen}(\delta\rightarrow\mathcal{L}_\zeta)
=2\zeta_\nu\check{X}^{\mu\nu}
-\nabla_\nu\check{Q}^{\mu\nu}}
\, , \label{ThetLieGFXQ0}
\ee
with $\check{Q}^{\mu\nu}={Q}^{\mu\nu}|_{{Z}\rightarrow\check{Z}}$.
Like the ${Q}^{\mu\nu}$ tensor, here the anti-symmetric tensor
$\check{Q}^{\mu\nu}$ is determined up to the divergence of an
arbitrary three-form as well. Equation
(\ref{ThetLieGFXQ0}) actually provides a direct way to verify that
the vector $\Theta^\mu_{gen}(\delta\rightarrow\mathcal{L}_\zeta)$
can be always decomposed into the form (\ref{ThetLieGF}). Because of
the allowance for the arbitrariness of the tensor
$\check{Z}_{(i)}^{\lambda_1\cdot\cdot\cdot\lambda_i\mu\nu\rho\sigma}$,
the decomposition for $\Theta^\mu_{gen}(\delta\rightarrow\mathcal{L}_\zeta)$
displayed by Eq. (\ref{ThetLieGFXQ0}) possesses an evident universality,
and it merely depends upon all the
$\check{Z}_{(i)}^{\lambda_1\cdot\cdot\cdot\lambda_i\mu\nu\rho\sigma}
\delta\nabla_{\lambda_1}
\cdot\cdot\cdot\nabla_{\lambda_i}R_{\mu\nu\rho\sigma}$ terms regardless
of whether or not such terms stem from the variation of the Lagrangian
$\sqrt{-g}{L}_{gen}$. In particular, within the situation where each
of the
$\check{Z}_{(i)}^{\lambda_1\cdot\cdot\cdot\lambda_i\mu\nu\rho\sigma}
\delta\nabla_{\lambda_1}
\cdot\cdot\cdot\nabla_{\lambda_i}R_{\mu\nu\rho\sigma}$ terms turns
into a component arising from the variation of the Lagrangian, and
the sum of all of them fulfills Eq. (\ref{VaricheckL}), in the
same spirit as the analysis performed towards the derivation of
Eq. (\ref{RelDivThLieVLag}) within Appendix \ref{appendB}, it is able
to be verified that equation (\ref{ThetLieGFXQ0}) can also appear as
a consequence out of the coexistence for both the two equations
(\ref{VaricheckL}) and (\ref{VarLaggen}) with $\delta$ substituted
by $\mathcal{L}_\zeta$. Additionally, within the same situation,
the rank-two tensor $\check{X}^{\mu\nu}$ is symmetric. Nevertheless,
the symmetry of this tensor can not be surely guaranteed in other
generic cases. What is more, by the aid of Eq. (\ref{ThetLieGFXQ0}),
equation (\ref{SumCheZiLieR}) is able to be interpreted as the direct
consequence from replacing $\delta$ in Eq. (\ref{DivThetLgen})
with $\mathcal{L}_\zeta$, which gives rise to
\be
\sum^{m}_{i=0}
\check{Z}_{(i)}^{\lambda_1\cdot\cdot\cdot\lambda_i\mu\nu\rho\sigma}
\mathcal{L}_\zeta\nabla_{\lambda_1}
\cdot\cdot\cdot\nabla_{\lambda_i}R_{\mu\nu\rho\sigma}
=\check{N}_{\mu\nu}\mathcal{L}_\zeta{g}^{\mu\nu}
+\nabla_\mu\Theta^\mu_{gen}(\delta\rightarrow\mathcal{L}_\zeta)
\, . \label{DivThetLgenLie}
\ee
In other words, for arbitrary tensors
$\check{Z}_{(i)}^{\lambda_1\cdot\cdot\cdot\lambda_i\mu\nu\rho\sigma}$s,
it is verified via direct calculations that
equation (\ref{DivThetLgen}) with $\delta$ transformed into
$\mathcal{L}_\zeta$ is valid identically.

On the basis of the surface term (\ref{ThetLgen}), making use of
the condition that equation (\ref{VaricheckL}) still holds after
the variation operator $\delta$ is replaced with the Lie derivative
$\mathcal{L}_\zeta$ along the arbitrary smooth vector field
$\zeta^\mu$, namely,
\be
\mathcal{L}_\zeta{L}_{gen}=\check{S}_{\mu\nu}
\mathcal{L}_\zeta{g}^{\mu\nu}+\sum^{m}_{i=0}
\check{Z}_{(i)}^{\lambda_1\cdot\cdot\cdot\lambda_i\mu\nu\rho\sigma}
\mathcal{L}_\zeta\nabla_{\lambda_1}
\cdot\cdot\cdot\nabla_{\lambda_i}{R}_{\mu\nu\rho\sigma}
\, , \label{LiedcheckL}
\ee
we can straightforwardly extend all the related results associated to the
Lagrangian (\ref{LagCovR}) to the corresponding ones for the Lagrangian
$\sqrt{-g}{L}_{gen}$ under the following substitution
\be
\frac{\partial{L}}{\partial{g}^{\mu\nu}}\rightarrow\check{S}_{\mu\nu}
\, , \qquad
\frac{\partial{L}}{\partial\nabla_{\lambda_1}\cdot\cdot\cdot
\nabla_{\lambda_i}{R}_{\mu\nu\rho\sigma}}
={Z}_{(i)}^{\lambda_1\cdot\cdot\cdot\lambda_i\mu\nu\rho\sigma}
\rightarrow
\check{Z}_{(i)}^{\lambda_1\cdot\cdot\cdot\lambda_i\mu\nu\rho\sigma}
\, . \label{SubsPgmnZ}
\ee
Particularly, equation (\ref{LiedcheckL}) suffices to lead to that
the expression ${E}^{\mu\nu}_{gen}$ for the field equations arising
from the variation of the Lagrangian (\ref{LagGen}) can be built
from the rank-two tensor $\check{X}^{\mu\nu}$ extracted out of
the surface term in the one-to-one manner, that is,
\be
{E}^{\mu\nu}_{gen}=\check{X}^{\mu\nu}
-\frac{1}{2}{L}_{gen}{g}^{\mu\nu}
\, . \label{EEomLgen}
\ee
Equation (\ref{LiedcheckL}) guarantees that the $\check{X}^{\mu\nu}$
tensor must be symmetric. In addition, it is worth mentioning that
Eq. (\ref{LiedcheckL}) is a sufficient and necessary condition for
Eq. (\ref{EEomLgen}) and the conservation law
$\nabla_\mu{E}^{\mu\nu}_{gen}=0$.

Apart from the aforementioned derivation for the expression
(\ref{EEomLgen}) relative to the equations of motion, with the help
of the Lie derivative of the Lagrangian density along the vector
field $\zeta^\mu$ presented by Eq. (\ref{LiedcheckL}), the scalar
$\sum^{m}_{i=0}
\check{Z}_{(i)}^{\lambda_1\cdot\cdot\cdot\lambda_i\mu\nu\rho\sigma}
\mathcal{L}_\zeta\nabla_{\lambda_1}
\cdot\cdot\cdot\nabla_{\lambda_i}{R}_{\mu\nu\rho\sigma}$ is also
able to produce such an expression. Concretely, when this scalar is
embedded into Eq. (\ref{LiedcheckL}),
yielding another pair 
\be
\big(\check{\mathcal{N}}^{\mu\nu},
\check{\mathcal{X}}^{\mu\nu}\big)
=\left(\frac{1}{2}{L}_{gen}{g}^{\mu\nu}-\check{S}^{\mu\nu},
\quad\frac{1}{2}{L}_{gen}{g}^{\mu\nu}\right)
\, , \label{PairNX2}
\ee 
by virtue of Eqs. (\ref{SumCheZiLieR2}) and (\ref{cheNXtrans}), 
the comparison between Eqs. (\ref{SumCheZiLieR}) and (\ref{LiedcheckL}) 
leads to that the rank-two symmetric and divergence-free tensor
$\Omega^{\mu\nu}=\check{N}^{\mu\nu}-\check{\mathcal{N}}^{\mu\nu}
=\check{X}^{\mu\nu}-\check{\mathcal{X}}^{\mu\nu}$ has to take the
specific value
\bea
\Omega^{\mu\nu}&=&\check{N}^{\mu\nu}+\check{S}^{\mu\nu}
-\frac{1}{2}{L}_{gen}{g}^{\mu\nu} \nn \\
&=&\check{X}^{\mu\nu}-\frac{1}{2}{L}_{gen}{g}^{\mu\nu}
\, . \label{OmegaLgen}
\eea
Consequently, we obtain the expression for the field equations
${E}^{\mu\nu}_{gen}=\Omega^{\mu\nu}$ together with
$\nabla_\mu{E}^{\mu\nu}_{gen}=0$ once again. This implies that
the $\Omega^{\mu\nu}$ tensor must convert to the expression for the
equations of motion if the scalar $\sum^{m}_{i=0}
\check{Z}_{(i)}^{\lambda_1\cdot\cdot\cdot\lambda_i\mu\nu\rho\sigma}
\mathcal{L}_\zeta\nabla_{\lambda_1}
\cdot\cdot\cdot\nabla_{\lambda_i}{R}_{\mu\nu\rho\sigma}$ is fully
determined by the Lie derivative of the Lagrangian density. Notably,
here this scalar is required to be put into the form (\ref{SumCheZiLieR})
in advance, which reserves the $\check{N}_{\mu\nu}$ tensor in the term
$\check{N}_{\mu\nu}\delta{g}^{\mu\nu}$ of Eq. (\ref{DivThetLgen}) and
takes the alternative form (\ref{DivThetLgenLie}). Within
Eq. (\ref{OmegaLgen}), the first equality represents nothing else but
the expression for the Euler-Lagrange equations of motion, while the
second equality denotes the expression of the field equations arising
from the surface term. As a result from Eq. (\ref{OmegaLgen}), it is
proved that both of the two expressions must coincide with each other
under the condition (\ref{LiedcheckL}). This further indicates that
the symmetric tensor $\check{S}^{\mu\nu}$ is in agreement with the
difference between both the tensors $\check{X}^{\mu\nu}$ and
$\check{N}^{\mu\nu}$, namely,
$\check{S}^{\mu\nu}=\check{X}^{\mu\nu}-\check{N}^{\mu\nu}$, which 
becomes the identity (\ref{Partgmnequal}) when the Lagrangian 
(\ref{LagGen}) is consistent with the one (\ref{LagCovR}).

Finally, based on the above results in this section, we present the
following \boxed{\textbf{conclusion}}: \\
\emph{For any scalar
$\sum^{m}_{i=0}
\check{Z}_{(i)}^{\lambda_1\cdot\cdot\cdot\lambda_i\mu\nu\rho\sigma}
\delta\nabla_{\lambda_1}
\cdot\cdot\cdot\nabla_{\lambda_i}{R}_{\mu\nu\rho\sigma}$, where the
nonnegative integer $m$ and each of the tensors
$\check{Z}_{(i)}^{\lambda_1\cdot\cdot\cdot\lambda_i\mu\nu\rho\sigma}$s
are allowed to be arbitrary, the surface term $\Theta^\mu_{gen}$
arising from such a scalar is presented by equation (\ref{ThetLgen}),
which can be always decomposed into the form (\ref{ThetLieGFXQ0})
in terms of the second-rank tensor $\check{X}^{\mu\nu}$ and the
anti-symmetric one $\check{Q}^{\mu\nu}$ under the transformation
$\delta\rightarrow\mathcal{L}_\zeta$. Particularly, when the scalar
$\sum^{m}_{i=0}
\check{Z}_{(i)}^{\lambda_1\cdot\cdot\cdot\lambda_i\mu\nu\rho\sigma}
\delta\nabla_{\lambda_1}
\cdot\cdot\cdot\nabla_{\lambda_i}{R}_{\mu\nu\rho\sigma}$ is embedded
in equation (\ref{VaricheckL}) representing the variation of the
Lagrangian density ${L}_{gen}$ within equation (\ref{LagGen}), the
$\check{X}^{\mu\nu}$ tensor resulting from the surface term must be
able to participate in the construction of the expression
${E}^{\mu\nu}_{gen}$ for the field equations corresponding to the
Lagrangian $\sqrt{-g}{L}_{gen}$ in such a way that
${E}^{\mu\nu}_{gen}=\check{X}^{\mu\nu}
-\frac{1}{2}{L}_{gen}{g}^{\mu\nu}$ under the condition that
equation (\ref{LiedcheckL}), i.e. equation (\ref{VaricheckL})
with $\delta$ substituted by $\mathcal{L}_\zeta$, is valid.
Besides, the anti-symmetric tensor $\check{Q}^{\mu\nu}$ can be
regarded as the Noether charge two-form associated to this
Lagrangian.}

\section{Conclusions and discussions}\label{eight}

Within the present paper, starting from the surface term (\ref{SurfT})
stemming from the variation equation (\ref{VarLag3}) for
the generic Lagrangian (\ref{LagCovR}) describing
diffeomorphism invariant gravity theories without matter fields,
it has been explicitly illustrated that this
surface term with the variation operator $\delta$ replaced by
the Lie derivative $\mathcal{L}_\zeta$ along an arbitrary smooth
vector field $\zeta^\mu$, denoted by
$\Theta^\mu(\delta\rightarrow\mathcal{L}_\zeta)$, can be put into
the form (\ref{ThetLieGF}), which is specific to the one given
by Eq. (\ref{ThetaLie}) for the Lagrangian (\ref{LagCovR}).
Such a form for the surface term
$\Theta^\mu(\delta\rightarrow\mathcal{L}_\zeta)$ consists of the
symmetric tensor $X^{\mu\nu}$ together with the anti-symmetric
one ${Q}^{\mu\nu}$. Subsequently, by utilizing a crucial
equation (\ref{Partgmnequal}), it has been
proved that the rank-two symmetric tensor $X^{\mu\nu}$ is
in agreement with the functional derivative of the Lagrangian
density ${L}$ with respect to the metric, that is,
${X}_{\mu\nu}=\frac{\delta{L}}{\delta{g}^{\mu\nu}}$. Apparently,
as a consequence of this equality, the expression
$\tilde{E}^{\mu\nu}={X}^{\mu\nu}-\frac{1}{2}{L}{g}^{\mu\nu}$
resulting from the surface term is completely in agreement with
the one ${E}^{\mu\nu}$ for the Euler-Lagrange equations of motion
given by Eq. (\ref{EulerLagEq}). On the other hand, it has been
illustrated that the relationship $\tilde{E}^{\mu\nu}={E}^{\mu\nu}$
can be employed to produce the off-shell Noether current
${J}^{\mu}$ in Eq. (\ref{ConsCurr}). Its corresponding Noether charge
two-form turns out to be the anti-symmetric tensor ${Q}^{\mu\nu}$.
Hence we conclude that the equations of motion and the Noether charge
can be produced in a uniform framework on the basis of the surface term,
that is,
\be
\boxed{{E}^{\mu\nu}={X}^{\mu\nu}-\frac{1}{2}{L}{g}^{\mu\nu}
\xleftarrow{{X}^{\mu\nu}}
\Theta^\mu(\delta\rightarrow\mathcal{L}_\zeta)
\text{ in Eq. }(\ref{ThetaLie})
\xrightarrow{{Q}^{\mu\nu}}
{J}^\mu=\nabla_\nu{Q}^{\mu\nu}}
\, . \nn
\ee
To see clearly the procedure for the derivation of the field equations,
the entire process of doing so is summarized as Eq. (\ref{ProceOne}).
In fact, this process can be thought of
as the inverse of the one given by Eq. (\ref{ProceTwo}), according
to which the equations of motion can be also derived out of
the surface term as is what has been demonstrated
within Ref. \cite{JJPEQM}. Through the combination of
Eq. (\ref{ProceOne}) with (\ref{ProceTwo}), the appearance of
the form (\ref{ThetaLie}) for the surface term, together with
either of the two equations (\ref{LieLagdens}) and (\ref{DivJE}),
guarantees that there exist some equivalence relationships relative
to the expressions for the field equations and the conserved current,
exhibited by Eqs. (\ref{FourEquiv}) and (\ref{EmnequlCC}).

In particular, we have illustrated that both the equations
(\ref{LieLagdens}) and (\ref{DivJE}), resulting from the substitution
of the variation operator by the Lie derivative along the
diffeomorphism generator within the variation equations of the
Lagrangian in two different forms, are equivalent to each other
according to Eqs. (\ref{LieLEJequiR}) and (\ref{LieDLtoThet}) or
according to the two equations (\ref{RelDivThLieVLag}) and
(\ref{EquiRTieLV}). Due to this, at the final stage to confirm that
both the expressions ${E}^{\mu\nu}$ and $\tilde{E}^{\mu\nu}$ for the
equations of motion coincide with each other, it has been demonstrated
that each of the equations (\ref{LieLagdens}) (or Eq. (\ref{Partgmnequal})
derived from it) and (\ref{DivJE}) can serve as the sufficient condition
for achieving this. However, by comparison, equation (\ref{DivJE}) is in
advantage by virtue of its conciseness. As a matter of fact, on the basis
of Eq. (\ref{LieLEJequiR}), we are able to arrive at the significant
conclusion that the validity for either of Eqs. (\ref{LieLagdens})
and (\ref{DivJE}) is a sufficient and necessary condition for that the
rank-two tensor ${X}^{\mu\nu}$ can be associated with the Euler-Lagrange
expression ${E}^{\mu\nu}$ for the field equations through the one-to-one
correspondence ${E}^{\mu\nu}={X}^{\mu\nu}-\frac{1}{2}{L}{g}^{\mu\nu}$
and the covariant divergence for ${E}^{\mu\nu}$ fulfills
$\nabla_\mu{E}^{\mu\nu}=0$ simultaneously. Inspired with this
conclusion, we further provide the generic procedure to derive the
expression for the equations of motion from the surface term
perspective within Section \ref{seven}.

Apart from both the two equations (\ref{LieLagdens}) and (\ref{DivJE}),
the structure (\ref{ThetLieGF}) possessed by the surface term
with the variation operator $\delta$ transformed into the Lie
derivative $\mathcal{L}_\zeta$ plays an indispensable role in the
construction of the expression for the field equations from the view
of the surface term. Due to this, within the framework of any
diffeomorphism invariant theory of gravity described by the Lagrangian
(\ref{LagGen}) depending upon the metric, its Riemann tensor and the
arbitrary order covariant derivatives of the Riemann tensor, it has
been verified through Eq. (\ref{ThetLieGFXQ0}) pertaining to the
surface term (\ref{ThetLgen}) that the surface term can be always
decomposed into the form (\ref{ThetLieGF}). That is to say, the surface
term with $\delta\rightarrow\mathcal{L}_\zeta$ must be able to supply
the rank-two symmetric tensor that is in agreement with the functional
derivative of the Lagrangian density with respect to the metric under
the condition (\ref{LiedcheckL}).

In terms of the results here together with the ones given by our
previous work \cite{JJPEQM}, we put forward that the expression for
the field equations associated to the diffeomorphism invariant
theories of gravity can be extracted out of the surface term.
To do so, according to Eq. (\ref{ProcST}) and the three steps for
the derivation of the field equations summarized at the beginning
of Section \ref{seven}, first, varying the Lagrangian density
yields the surface term. Next, under the substitution
$\delta\rightarrow\mathcal{L}_\zeta$, the surface term is written
as the form (\ref{ThetLieGF}). Finally, due to the conclusion that
the second-rank tensor contracted with the vector field $\zeta^\mu$
in Eq. (\ref{ThetLieGF}) is identified with the functional derivative
of the Lagrangian density $L$ with regard
to the metric, the expression for the field equations turns out
to be this second-rank tensor minus $\frac{1}{2}{L}{g}^{\mu\nu}$.
By contrast with the conventional approach to express the field
equations as the Euler-Lagrange equations through the variation
of the Lagrangian, resorting to the surface term for producing
equations of motion at least has the following merits. First,
the derivative of the Lagrangian density with respect to the
metric is completely excluded throughout the entire process for
the derivation of the field equations. This simplifies the field
equations greatly. Second, on the basis of Eq. (\ref{EmnequlCC})
(see Fig. \ref{FigEJ} as well), we find that the surface term
provides the advantage of revealing the underlying connections
between the conserved current and the expression for the equations
of motion by establishing their equivalence relationships in
several ways. This straightforwardly supports that the
diffeomorphism invariance symmetry of the Lagrangian can determine
the equations of motion for the gravitational field in addition to
give rise to the conserved current. Third, as what has been shown,
the surface term with the variation operator transforming into the Lie
derivative can be expressed as the form encompassing a symmetric
tensor and an anti-symmetric one, corresponding to the
functional derivative of the Lagrangian density with respect to
the metric and the Noether charge two-form respectively. Due to
this, the surface term offers a unified description for the field
equations and the Noether charge. Therefore, apart from the
former, the latter can be also derived at the same time. Fourth,
the expression for the field equations arising from the surface
term provides great convenience for calculating its covariant
divergence. Fifth, during the process to derive the field
equations, we obtain the explicit expressions for the surface
term and the Noether charge two-form simultaneously. These
quantities can be directly employed to define the conserved
charges of the gravity theories.

Within the context of the pure gravity theories
admitting diffeomorphism invariance symmetry, according to
Eq. (\ref{EmnequlCC}) or (\ref{DivEgenEoM}), we obtain the
conclusion that the expression for the field equations can be
derived out of the vanishing of its covariant divergence.
Besides, according to Eq. (\ref{EmnequlCC}) or (\ref{DivJgenEoM}),
the expression for the field equations can also result from
the disappearance of the divergence for the off-shell Noether current.

What is more, we have demonstrated that the expression
$\tilde{E}^{\mu\nu}$ for the field equations arising from the
surface term can be regarded as the most natural generalization
of the one associated to the theories of gravity only depending
upon the metric and the Riemann tensor given by Ref. \cite{Pady}
to the
higher derivative theories of gravity in the presence of the
arbitrary order covariant derivatives of the Riemann tensor.
To this point, our procedure for the
extraction of the field equations out of the surface term offers an
alternative path towards the reproduction and the generalization for
the field equations through the method proposed by Padmanabhan in
Ref. \cite{Pady}. Inspired with this, we have further revealed
that Eq. (\ref{EquiReLieLE}) or the first equivalence relationship
in Eq. (\ref{LieLEJequiR}) encompasses the main implications of
Padmanabhan's method. According to Eqs. (\ref{ProcST}) and
(\ref{LieLEJequiR}), although Padmanabhan's method and our approach
from the perspective of surface term produce the same equations of
motion, there exist apparent differences between both of them.
In fact, within the context of Padmanabhan's method, the concrete
Euler-Lagrange equations of motion arising from the variation of the
Lagrangian has to be supplied and it is unavoidable to deal with
the Lie derivative of the Lagrangian density to get an equality
associated to its derivative with respect to the metric tensor.
In contrast, our approach completely abandons the above, and it only
involves the handling of the surface term derived out of the
variation of the Lagrangian density.

\section*{Acknowledgments}

This work was supported by the National Natural Science
Foundation of China under Grant No. 12565009.

\appendix
\section{The structure of the surface term with the variation
operator substituted by the Lie derivative}
\label{appendA}

\subsection{General formalism}\label{appendAs1}

The decomposition of the surface term under the transformation
$\delta\rightarrow\mathcal{L}_\zeta$ plays a fundamental role in
deriving our main results. The desired decomposition form for
this quantity gives rise to two tensors corresponding
to the expression for equations of motion and the Noether charge
two-form respectively. Consequently, for the sake of importance and
completeness, here we give some discussions on the structure of
the surface term by partially following our earlier work \cite{JJPEQM}.

Without loss of generality, focusing on the generic Lagrangian
(\ref{LagGen}) accommodating diffeomorphism invariance symmetry,
that is, $\sqrt{-g}{L}_{gen}=\sqrt{-g}{L}_{gen}\left({g}^{\mu\nu},
{R}_{\alpha\beta\rho\sigma},\nabla_\lambda\right)$, we vary this
Lagrangian with respect to the metric, leading to
\cite{IyerWald,LeeWald,NoeChWald}
\be
\delta\big(\sqrt{-g}{L}_{gen}\big)
=\sqrt{-g}\left(-{E}^{\mu\nu}_{gen}\delta{g}_{\mu\nu}
+\nabla_\mu\Theta^\mu_{gen}\right)
\, . \label{VarLaggen}
\ee
As usual, the symmetric rank-two tensor ${E}^{\mu\nu}_{gen}$ in
Eq. (\ref{VarLaggen}) represents the expression for the
Euler-Lagrange equations of motion. For the purpose in this work,
as well as in our previous work \cite{JJPEQM}, the surface term
$\Theta^\mu_{gen}$ with $\delta$ replaced by $\mathcal{L}_\zeta$
is supposed to be decomposed into the following general form
\be
\Theta^\mu_{gen}(\delta\rightarrow\mathcal{L}_\zeta)
=2\zeta_\nu(\text{rank-2 tensor})^{\mu\nu}
-\nabla_\nu(\text{\textbf{anti-symmetric tensor}})^{\mu\nu}
\, . \label{ThetLieGF}
\ee
Here the first second-rank tensor at the right hand side of
the above equation, denoted by $(\text{rank-2 tensor})^{\mu\nu}$,
is required to be independent of the vector field $\zeta^\mu$
attributed to the fact that the dependence of $\Theta^\mu_{gen}$ on
$\delta{g}^{\mu\nu}$ and the variation of the covariant derivatives
acting on ${R}_{\alpha\beta\rho\sigma}$ is linear.
That is to say, it is assumed that there exists the rank-two
tensor $(\text{rank-2 tensor})^{\mu\nu}$ irrelevant to $\zeta^\mu$,
satisfying the condition that the difference between
$2\zeta_\nu(\text{rank-2 tensor})^{\mu\nu}$ and
$\Theta^\mu_{gen}(\delta\rightarrow\mathcal{L}_\zeta)$ is
divergence-free. For the sake of providing some clues for achieving
the above decomposition for
$\Theta^\mu_{gen}(\delta\rightarrow\mathcal{L}_\zeta)$, we find
that equation (\ref{ThetLieGF}) is tightly connected to the
covariant divergence of the expression ${E}^{\mu\nu}_{gen}$ for
equations of motion. In fact, under the condition that the variation
equation (\ref{VarLaggen}) for the Lagrangian is still valid after
the operator $\delta$ is replaced with the Lie derivative
$\mathcal{L}_\zeta$, the existence of the form (\ref{ThetLieGF})
can be completely guaranteed as long as there exists the generalized
Bianchi identity
\be
\nabla_\mu{E}^{\mu\nu}_{gen}=0
\, . \label{VanDivEgen}
\ee
To see this, taking into consideration of the situation where the
variation of the Lagrangian is carried out through the diffeomorphism
generated by the arbitrary vector field $\zeta^\mu$, one is able to
replace the variation operator $\delta$ with the Lie derivative
$\mathcal{L}_\zeta$. Doing so gives rise to the following identity
\be
\nabla_\mu\left[2\zeta_\nu\left({E}^{\mu\nu}_{gen}
+\frac{1}{2}{L}_{gen}{g}^{\mu\nu}\right)
-\Theta^\mu_{gen}(\delta\rightarrow\mathcal{L}_\zeta)\right]
=2\zeta_\nu\nabla_\mu{E}^{\mu\nu}_{gen}
\, . \label{VarLaggenLie}
\ee
In accordance with the spirit of the work \cite{JJPEQM}, here we
stress that the substitution $\delta\rightarrow\mathcal{L}_\zeta$
performed towards the variation equation (\ref{VarLaggen}) of the
Lagrangian is supposed to be allowed so that Eq. (\ref{VarLaggenLie})
holds identically throughout the present appendix. The identity
(\ref{VarLaggenLie}) is instrumental in deriving the main results
in this appendix. Owing to the disappearance of the right hand side
of Eq. (\ref{VarLaggenLie}), then utilizing Poincare lemma gives
rise to
\be
\Theta^\mu_{gen}(\delta\rightarrow\mathcal{L}_\zeta)
=2\zeta_\nu(\text{\textbf{symmetric tensor}})^{\mu\nu}
-\nabla_\nu(\text{\textbf{anti-symmetric tensor}})^{\mu\nu}
\, , \label{ThetLieGF2}
\ee
where the rank-two symmetric tensor contracted with the vector
field $\zeta^\mu$ takes the specific expression:
$(\text{\textbf{symmetric tensor}})^{\mu\nu}=
{E}^{\mu\nu}_{gen}+\frac{1}{2}{L}_{gen}{g}^{\mu\nu}$. Thus,
on the basis of Eqs. (\ref{VanDivEgen}) and (\ref{ThetLieGF2}),
one is able to conclude that the vanishing of the divergence
of the expression for the field equations is a sufficient
condition for that
$\Theta^\mu_{gen}(\delta\rightarrow\mathcal{L}_\zeta)$
bears the same structure as that of Eq. (\ref{ThetLieGF}).

In light of Eq. (\ref{ThetLieGF2}), we further arrive at the stronger
conclusion that the surface term $\Theta^\mu_{gen}$
under the replacement $\delta\rightarrow\mathcal{L}_\zeta$ can be
decomposed into the linear combination for the divergence of a
second-rank anti-symmetric tensor together with the contraction
between the vector field $\zeta^\mu$ and another second-rank
symmetric tensor. However, if $\Theta^\mu_{gen}$ is supposed to
be decomposed as $\Theta^\mu_{gen}=\sum_{k=1}\Theta^\mu_{(k)}$,
it is unnecessary to require each component
$\Theta^\mu_{(k)}(\delta\rightarrow\mathcal{L}_\zeta)$
to possess the same structure as the one of Eq. (\ref{ThetLieGF2}).

Apart from the aforementioned consequence that equation
(\ref{ThetLieGF}) can be derived out of the condition
$\nabla_\mu{E}^{\mu\nu}_{gen}=0$, there exists the fact that its inverse
holds true as well, that is, the covariant divergence of the
expression for the field equations vanishes identically if the
surface term under the substitution
$\delta\rightarrow\mathcal{L}_\zeta$ can be written as the form
(\ref{ThetLieGF}). To do so, according to Eq. (\ref{ThetLieGF}),
we perform a decomposition of the surface term
$\Theta^\mu_{gen}(\delta\rightarrow\mathcal{L}_\zeta)$ into
two components. One of them consists of the contraction between
the vector field $\zeta^\mu$ and the rank-two tensor
${X}^{\mu\nu}_{gen}$, which is irrelevant to the former, and the
other is the divergence for the anti-symmetric rank-two tensor
${Q}^{\mu\nu}_{gen}$. Specifically,
$\Theta^\mu_{gen}(\delta\rightarrow\mathcal{L}_\zeta)$
exhibits the form
\be
\Theta^\mu_{gen}(\delta\rightarrow\mathcal{L}_\zeta)
=2\zeta_\nu{X}^{\mu\nu}_{gen}
-\nabla_\nu{Q}^{\mu\nu}_{gen}
\, . \label{ThetLieGFXQ}
\ee
Apparently, by comparison between Eqs. (\ref{ThetLieGFXQ0}) and
(\ref{ThetLieGFXQ}), it is easy to see that both the two rank-two tensors
${X}^{\mu\nu}_{gen}$ and ${Q}^{\mu\nu}_{gen}$ given by the above equation
can be specific to ${X}^{\mu\nu}_{gen}=\check{X}^{\mu\nu}$ and
${Q}^{\mu\nu}_{gen}=\check{Q}^{\mu\nu}$, respectively. After substituting
Eq. (\ref{ThetLieGFXQ}) into (\ref{VarLaggenLie}), we have
\be
\zeta_\nu\nabla_\mu\left({X}^{\mu\nu}_{gen}
-\frac{1}{2}{L}_{gen}{g}^{\mu\nu}\right)+
\left({X}^{\mu\nu}_{gen}
-{E}^{\mu\nu}_{gen}-\frac{1}{2}{L}_{gen}{g}^{\mu\nu}\right)
\nabla_\mu\zeta_\nu
=0
\, . \label{VarLaggenLieXQ}
\ee
Here equation (\ref{VarLaggenLieXQ}) holds for any vector field
$\zeta^\mu$. Consequently, as is what has been shown in
Ref. \cite{JJPEQM}, the expression ${E}^{\mu\nu}_{gen}$ for the
equations of motion can be reproduced by means of
Eq. (\ref{VarLaggenLieXQ}), that is,
\be
{E}^{\mu\nu}_{gen}={X}^{\mu\nu}_{gen}
-\frac{1}{2}{L}_{gen}{g}^{\mu\nu}
\, . \label{Emngen}
\ee
Apart from this, equation (\ref{VarLaggenLieXQ}) enables one to write
down the following identity
\be
\nabla_\mu\left({X}^{\mu\nu}_{gen}
-\frac{1}{2}{L}_{gen}{g}^{\mu\nu}\right)
=\nabla_\mu{E}^{\mu\nu}_{gen}=0
\, . \label{CoDivEmnXQ}
\ee
Due to the above, one observes that equation (\ref{VanDivEgen})
is a necessary and sufficient condition for the existence of the
form (\ref{ThetLieGF}) in which the surface term
$\Theta^\mu_{gen}(\delta\rightarrow\mathcal{L}_\zeta)$ can be, namely,
\be
\nabla_\mu{E}^{\mu\nu}_{gen}=0\quad\Leftrightarrow\quad
\text{Equation } (\ref{ThetLieGF})
\, . \label{NSconTheLie}
\ee
Here equation (\ref{NSconTheLie}) provides another way to verify
the disappearance of the covariant divergence for the expression
of the field equations, that is, it is only required to confirm
that $\Theta^\mu_{gen}(\delta\rightarrow\mathcal{L}_\zeta)$ can
be decomposed into the form (\ref{ThetLieGF}). If this fails,
the generalized
Bianchi identity $\nabla_\mu{E}^{\mu\nu}_{gen}=0$ must break down.
As a matter of fact, for the diffeomorphism invariant Lagrangian
(\ref{LagGen}), repeating the related
analysis for the Lagrangian (\ref{LagCovR}), we find that the
surface term $\Theta^\mu_{gen}(\delta\rightarrow\mathcal{L}_\zeta)$
is able to be expressed as the form (\ref{ThetLieGF}). Consequently,
$\nabla_\mu{E}^{\mu\nu}_{gen}=0$ holds identically. Then identity
(\ref{VarLaggenLie}) turns into
\be
\nabla_\mu{J}^\mu_{gen}=0
\, , \label{ConEqJgen}
\ee
in which the conserved current ${J}^\mu_{gen}$ is defined by
\be
{J}^\mu_{gen}=2\zeta_\nu{E}^{\mu\nu}_{gen}
+\zeta^\mu{L}_{gen}
-\Theta^\mu_{gen}(\delta\rightarrow\mathcal{L}_\zeta)
\, . \label{ConCuJgen}
\ee
According to the connection between the divergences for
${J}^{\mu}_{gen}$ and ${E}^{\mu\nu}_{gen}$
given by the identity (\ref{VarLaggenLie}), equation
(\ref{NSconTheLie}) is enhanced as
\be
\nabla_\mu{E}^{\mu\nu}_{gen}=0\quad\Leftrightarrow\quad
\text{Equation } (\ref{ThetLieGF})\quad\Leftrightarrow\quad
\nabla_\mu{J}^{\mu}_{gen}=0
\, . \label{EJgenTheLie}
\ee
In addition to this, attributed to the
fact that $\nabla_\mu{E}^{\mu\nu}_{gen}=0$ guarantees the
decomposition for
$\Theta^\mu_{gen}(\delta\rightarrow\mathcal{L}_\zeta)$ given by
Eq. (\ref{ThetLieGFXQ}) and the ${X}^{\mu\nu}_{gen}$ tensor
determined by this equation is in one-to-one correspondence
to ${E}^{\mu\nu}_{gen}$ displayed by Eq. (\ref{Emngen}), one
arrives at an interesting conclusion
\be
\nabla_\mu{E}^{\mu\nu}_{gen}=0\quad\Rightarrow\quad
{E}^{\mu\nu}_{gen}={X}^{\mu\nu}_{gen}
-\frac{1}{2}{L}_{gen}{g}^{\mu\nu}
\, . \label{DivEgenEoM}
\ee
This indicates that the expression for the field equations even can
arise from the condition that it is divergence-free. Moreover,
equation (\ref{EJgenTheLie}) shows that Eq. (\ref{DivEgenEoM})
is equivalent to
\be
\nabla_\mu{J}^{\mu}_{gen}=0\quad\Rightarrow\quad
{E}^{\mu\nu}_{gen}={X}^{\mu\nu}_{gen}
-\frac{1}{2}{L}_{gen}{g}^{\mu\nu}
\, . \label{DivJgenEoM}
\ee
That is to say, the expression for the field equations can be
also derived out of the divergence-free equation for the
conserved current.

With Eq. (\ref{ThetLieGFXQ}) in hand, we turn our attention to
proving that the second-rank tensor contracted with the vector
field $\zeta^\mu$ is unique as long as
the surface term $\Theta^\mu_{gen}$ under transformation
$\delta\rightarrow\mathcal{L}_\zeta$ can be expressed as the
form (\ref{ThetLieGF}). That is to say, once
$\Theta^\mu_{gen}(\delta\rightarrow\mathcal{L}_\zeta)$ takes
on the structure displayed by Eq. (\ref{ThetLieGF}), the
$X^{\mu\nu}_{gen}$ tensor is uniquely determined. To see this,
apart from Eq. (\ref{ThetLieGFXQ}), it is assumed that
$\Theta^\mu_{gen}(\delta\rightarrow\mathcal{L}_\zeta)$
has another similar form
\be
\Theta^\mu_{gen}(\delta\rightarrow\mathcal{L}_\zeta)
=2\zeta_\nu\hat{X}^{\mu\nu}_{gen}
-\nabla_\nu\hat{Q}^{\mu\nu}_{gen}
\, . \label{ThetaLieAF}
\ee
As before, here the second-rank tensor ${X}^{\mu\nu}_{gen}$ bears
no relation to the vector field $\zeta^\mu$, and
$\hat{Q}^{\mu\nu}_{gen}$ represents a certain anti-symmetric tensor.
Computing the divergence of the difference between
Eqs. (\ref{ThetLieGFXQ}) and (\ref{ThetaLieAF}), we get
\be
2\zeta_\nu\nabla_\mu\Big({X}^{\mu\nu}_{gen}-\hat{X}^{\mu\nu}_{gen}\Big)
+2\Big({X}^{\mu\nu}_{gen}-\hat{X}^{\mu\nu}_{gen}\Big)\nabla_\mu\zeta_\nu
=\nabla_\mu\nabla_\nu\Big({Q}^{\mu\nu}_{gen}-\hat{Q}^{\mu\nu}_{gen}\Big)
=0
\, . \label{DivdifX}
\ee
Thus, for the validity of the above equation for any vector field
$\zeta^\mu$, we further have
\be
\hat{X}^{\mu\nu}_{gen}={X}^{\mu\nu}_{gen}
\, . \label{EquaXhatX}
\ee
This implies that the ${X}^{\mu\nu}_{gen}$ tensor is unique.
As a result of Eq. (\ref{Emngen}), it has to be symmetric
and takes the specific form
\be
{X}^{\mu\nu}_{gen}={g}^{\mu\rho}{g}^{\nu\rho}
\frac{\delta{L}_{gen}}{\delta{g}^{\rho\sigma}}
=-\frac{\delta{L}_{gen}}{\delta{g}_{\mu\nu}}
\, . \label{XgenindelL}
\ee
By the combination of Eqs. (\ref{ThetLieGF}) with (\ref{ThetLieGFXQ}),
equation (\ref{NSconTheLie}) can be further recast into
\be
\nabla_\mu{E}^{\mu\nu}_{gen}=0\Leftrightarrow
\Theta^\mu_{gen}(\delta\rightarrow\mathcal{L}_\zeta)
=-2\zeta_\nu\frac{\delta{L}_{gen}}{\delta{g}_{\mu\nu}}
-\nabla_\nu(\text{\textbf{anti-symmetric tensor}})^{\mu\nu}
\, . \label{NSconTheLie2}
\ee

Although it is proved that the $X^{\mu\nu}_{gen}$ tensor is unique,
this does not hold true for the anti-symmetric tensor
$Q^{\mu\nu}_{gen}$. In fact, the $Q^{\mu\nu}_{gen}$ tensor is
determined up to the divergence of an arbitrary three-form. However,
the covariant divergence of the rank-two anti-symmetric tensor
remains the same, that is,
\be
\nabla_\nu(\text{\textbf{anti-symmetric tensor}})^{\mu\nu}=
\nabla_\nu{Q}^{\mu\nu}_{gen}
\, . \label{CovDivQgen}
\ee
What is more, due to the fact that the expression for field equations
${E}^{\mu\nu}_{gen}=X^{\mu\nu}_{gen}-\frac{1}{2}{L}_{gen}{g}^{\mu\nu}$,
the uniqueness of the $X^{\mu\nu}_{gen}$ tensor further determines
that ${E}^{\mu\nu}_{gen}$ is unique as well. As a consequence,
one arrives at the conclusion that every Lagrangian corresponds
to the unique field equations.

\subsection{The application in Lagrangian (\ref{LagCovR})}
\label{appendAs2}

Let us pay attention to the specific situation related to the
Lagrangian (\ref{LagCovR}). As is shown within Section \ref{two},
generally, the surface term $\Theta^\mu$ resulting from the variation
of the Lagrangian (\ref{LagCovR}) can be split into $\Theta^\mu_{P}$ and
$\bar{\Theta}^\mu_{(i)}$s $(i=1,2,\cdot\cdot\cdot,m)$.
If each of them under the transformation
$\delta\rightarrow\mathcal{L}_\zeta$ is able to be expressed as
the form (\ref{ThetLieGF}), their sum naturally inherits such a form.
In fact, for $\Theta^\mu_{P}$ coming from the scalar
${P}^{\mu\nu\rho\sigma}\delta{R}_{\mu\nu\rho\sigma}$ in the variation
of the Lagrangian, the form (\ref{ThetPLie}) for
$\Theta^\mu_{P}(\delta\rightarrow\mathcal{L}_\zeta)$ obviously belongs
to the one in Eq. (\ref{ThetLieGF}). On the other hand, for the vector
$\bar{\Theta}^\mu_{(i)}$, the definition of the Lie derivative for
tensors ensures that the surface term $\bar{\Theta}^\mu_{(i)}$
with $\delta\rightarrow\mathcal{L}_\zeta$ always behaves as
\bea
\bar{\Theta}^\mu_{(i)}(\delta\rightarrow\mathcal{L}_\zeta)&=&
\zeta_\nu\Phi^{\mu\nu}_{(i)}
+\Psi^{\lambda\mu\nu}_{(i)}\nabla_\nu\zeta_\lambda \nn \\
&=&\zeta_\nu\left(\Phi^{\mu\nu}_{(i)}
-\nabla_\lambda\Psi^{\nu\mu\lambda}_{(i)}
\right) +\nabla_\nu\big(\zeta_\lambda\Psi^{\lambda\mu\nu}_{(i)}\big)
\, , \label{ThetiLieGen}
\eea
where the tensors $\Phi^{\mu\nu}_{(i)}$ and $\Psi^{\lambda\mu\nu}_{(i)}$
are irrelevant to the vector field $\zeta^\mu$.
From Eq. (\ref{ThetiLieGen}), one observes that the surface term
$\bar{\Theta}^\mu_{(i)}(\delta\rightarrow\mathcal{L}_\zeta)$
can be put into the form (\ref{ThetLieGF}) if
\be
\Psi^{\lambda\mu\nu}_{(i)}=\Psi^{\lambda[\mu\nu]}_{(i)}
\, . \label{PsiAntiSym}
\ee
Specifically, in terms of Eq. (\ref{Thetbar}), by making use of
the identity
\bea
&&\sum^i_{k=1}(-1)^{k-1}\Big(\nabla_{\lambda_{k-1}}\cdot\cdot\cdot
\nabla_{\lambda_{1}}
{Z}^{\lambda_1\cdot\cdot\cdot\lambda_{k-1}\mu
\lambda_{k+1}\cdot\cdot\cdot\lambda_{i}\alpha\beta\rho\sigma}_{(i)}
\Big)\mathcal{L}_\zeta\nabla_{\lambda_{k+1}}
\cdot\cdot\cdot\nabla_{\lambda_{i}}{R}_{\alpha\beta\rho\sigma}\nn \\
&=&\zeta_\nu
\sum^i_{k=1}(-1)^{k-1}\Big(\nabla_{\lambda_{k-1}}\cdot\cdot\cdot
\nabla_{\lambda_{1}}Z^{\lambda_1\cdot\cdot\cdot\lambda_{k-1}\mu
\lambda_{k+1}\cdot\cdot\cdot\lambda_{i}\alpha\beta\rho\sigma}_{(i)}
\Big) \nabla^\nu \nabla_{\lambda_{k+1}}
\cdot\cdot\cdot\nabla_{\lambda_{i}}R_{\alpha\beta\rho\sigma} \nn \\
&&-{U}^{\lambda\mu\nu}_{(i)}\nabla_\nu\zeta_\lambda
\, , \label{deltoLienabR}
\eea
the second-rank tensor $\Phi^{\mu\nu}_{(i)}$ is read off as
\be
\Phi^{\mu\nu}_{(i)}=
\sum^i_{k=1}(-1)^{k-1}\Big(\nabla_{\lambda_{k-1}}\cdot\cdot\cdot
\nabla_{\lambda_{1}}Z^{\lambda_1\cdot\cdot\cdot\lambda_{k-1}\mu
\lambda_{k+1}\cdot\cdot\cdot\lambda_{i}\alpha\beta\rho\sigma}_{(i)}
\Big) \nabla^\nu \nabla_{\lambda_{k+1}}
\cdot\cdot\cdot\nabla_{\lambda_{i}}R_{\alpha\beta\rho\sigma}
\, , \label{Phimnidef}
\ee
and the third-rank tensor $\Psi^{\lambda\mu\nu}_{(i)}$ has the form
\bea
\Psi^{\lambda\mu\nu}_{(i)}&=&-{U}^{\lambda\mu\nu}_{(i)}
+{U}^{(\nu|\mu|\lambda)}_{(i)}
+{U}^{(\nu\lambda)\mu}_{(i)}
-{U}^{\mu(\nu\lambda)}_{(i)} \nn \\
&=&-\left({U}^{\lambda[\mu\nu]}_{(i)}
+{U}^{[\mu\nu]\lambda}_{(i)}
+{U}^{[\mu|\lambda|\nu]}_{(i)}\right)
\, . \label{Psimnidef}
\eea
Obviously, $\Psi^{\lambda\mu\nu}_{(i)}=\Psi^{\lambda[\mu\nu]}_{(i)}$
naturally obeys the constraint (\ref{PsiAntiSym}).
As a consequence, the vector field
$\bar{\Theta}^\mu_{(i)}(\delta\rightarrow\mathcal{L}_\zeta)$ is able
to be expressed as the form (\ref{ThetLieGF}).

Therefore, with the condition (\ref{PsiAntiSym}) satisfied, the
surface term $\Theta^\mu(\delta\rightarrow\mathcal{L}_\zeta)$
can be written as the form (\ref{ThetaLie}), in which the tensors
$X^{\mu\nu}$ and $Q^{\mu\nu}$ are presented by
\bea
X^{\mu\nu}&=&{P}^{\mu\lambda\rho\sigma}{R}^{\nu}_{~\lambda\rho\sigma}
-2\nabla_{\rho}\nabla_{\sigma}{P}^{\rho\mu\nu\sigma}
+\frac{1}{2}\sum^m_{i=1}
\big(\Phi^{\mu\nu}_{(i)}-\nabla_\lambda\Psi^{\nu\mu\lambda}_{(i)}\big)
\, , \label{GenXdef} \\
Q^{\mu\nu}&=&2{P}^{\mu\nu\rho\sigma}
\nabla_{\rho}\zeta_{\sigma}
+4\zeta_\rho\nabla_\sigma {P}^{\mu\nu\rho\sigma}
-6{P}^{\mu[\nu\rho\sigma]}\nabla_\rho\zeta_\sigma
-\zeta_\lambda\sum^m_{i=1}\Psi^{\lambda\mu\nu}_{(i)}
\, , \label{GenQdef}
\eea
respectively. Due to the above, in order to guarantee that the
surface term $\Theta^\mu(\delta\rightarrow\mathcal{L}_\zeta)$
possesses the form (\ref{ThetLieGF}), the sole constraint on
this term is to confirm that the third-rank tensor
$\Psi^{\lambda\mu\nu}_{(i)}$ fulfills the condition (\ref{PsiAntiSym}).

\section{The coexistence of equations (\ref{LieLagdens}) and
(\ref{DivJE}) arising from the equivalence between equations
(\ref{VarLag}) and (\ref{VarLag3})}\label{appendB}

Both the two equations (\ref{VarLag}) and (\ref{VarLag3}) are two
different but equivalent forms for the variation of the Lagrangian
(\ref{LagCovR}). They give rise to Eqs. (\ref{LieLagdens}) and
(\ref{DivJE}) under the substitution
$\delta\rightarrow\mathcal{L}_\zeta$, respectively. In the present
appendix, we will demonstrate that Eqs. (\ref{LieLagdens}) and
(\ref{DivJE}) always coexist through straightforward calculations.

For convenience, we introduce a scalar ${V}$ defined through
\be
{V}=\sum^{m}_{i=0}
{Z}_{(i)}^{\lambda_1\cdot\cdot\cdot\lambda_i\mu\nu\rho\sigma}
\mathcal{L}_\zeta\nabla_{\lambda_1}
\cdot\cdot\cdot\nabla_{\lambda_i}{R}_{\mu\nu\rho\sigma}
-{P}^{\alpha\beta\rho\sigma}\mathcal{L}_\zeta{R}_{\alpha\beta\rho\sigma}
+{B}^{\mu\nu}\mathcal{L}_\zeta{g}_{\mu\nu}
\, . \label{ScalVdef}
\ee
By the aid of the two identities (\ref{PWBexpan}) and (\ref{DivWmn}),
${V}$ is written as
\bea
{V}&=&2{W}^{\mu\nu}\nabla_\mu\zeta_\nu
+\zeta_\nu\left(\sum^{m}_{i=0}
{Z}_{(i)}^{\lambda_1\cdot\cdot\cdot\lambda_i\alpha\beta\rho\sigma}
\nabla^\nu\nabla_{\lambda_1}
\cdot\cdot\cdot\nabla_{\lambda_i}{R}_{\alpha\beta\rho\sigma}
-{P}^{\alpha\beta\rho\sigma}\nabla^\nu{R}_{\alpha\beta\rho\sigma}\right)
\nn \\
&=&2{W}^{\mu\nu}\nabla_\mu\zeta_\nu+2\zeta_\nu\nabla_\mu{W}^{\mu\nu}\nn \\
&=&2\nabla_\mu\left(\zeta_\nu{W}^{\mu\nu}\right)
\, . \label{ScalVdef2}
\eea
This implies that the scalar ${V}$ is a total divergence term.
Through the combination of Eq. (\ref{ScalVdef2}) with
$2\nabla_\mu\big(\zeta_\nu{W}^{\mu\nu}\big)
=\sum^{m}_{i=0}\nabla_\mu\bar{\Theta}^\mu_{(i)}
(\delta\rightarrow\mathcal{L}_\zeta)$ stemming from
Eq. (\ref{BarThetLie}), we gain the following identity
\be
{V}=\nabla_\mu\left(\sum^{m}_{i=0}\bar{\Theta}^\mu_{(i)}
(\delta\rightarrow\mathcal{L}_\zeta)\right)
\, . \label{IdentVTheti}
\ee
On the other hand, by utilizing the two identities
(\ref{PDivRiem}) and (\ref{Divdel2Pijp}), we have the identity
\bea
&&{P}^{\mu\nu\rho\sigma}\mathcal{L}_\zeta{R}_{\mu\nu\rho\sigma}
-\big({P}^{\mu\lambda\rho\sigma}{R}^\nu_{~\lambda\rho\sigma}
+2\nabla_\rho\nabla_\sigma{P}^{\rho\mu\nu\sigma}\big)
\mathcal{L}_\zeta{g}_{\mu\nu} \nn \\
&&=2\nabla_\mu\left[\zeta_\nu\big(
{P}^{\mu\lambda\rho\sigma}{R}^\nu_{~\lambda\rho\sigma}
-2\nabla_\rho\nabla_\sigma{P}^{\rho\mu\nu\sigma}\big)\right]
\, . \label{PLieRLieg}
\eea
With the help of Eq. (\ref{ThetPLie}), the identity (\ref{PLieRLieg})
is expressed in terms of the covariant divergence of
$\Theta^\mu_{P}(\delta\rightarrow\mathcal{L}_\zeta)$ through
\be
\nabla_\mu\Theta^\mu_{P}(\delta\rightarrow\mathcal{L}_\zeta)
={P}^{\mu\nu\rho\sigma}\mathcal{L}_\zeta{R}_{\mu\nu\rho\sigma}
-\big({P}^{\mu\lambda\rho\sigma}{R}^\nu_{~\lambda\rho\sigma}
+2\nabla_\rho\nabla_\sigma{P}^{\rho\mu\nu\sigma}\big)
\mathcal{L}_\zeta{g}_{\mu\nu}
\, . \label{PLieRLieg2}
\ee
Subsequently, by means of substituting the identity (\ref{PLieRLieg})
into the one (\ref{ScalVdef2}) to eliminate the scalar
${P}^{\mu\nu\rho\sigma}\mathcal{L}_\zeta{R}_{\mu\nu\rho\sigma}$
within the latter, we obtain another identity
\bea
&&\sum^{m}_{i=0}
{Z}_{(i)}^{\lambda_1\cdot\cdot\cdot\lambda_i\mu\nu\rho\sigma}
\mathcal{L}_\zeta\nabla_{\lambda_1}
\cdot\cdot\cdot\nabla_{\lambda_i}{R}_{\mu\nu\rho\sigma}= \nn \\
&&-\big({B}^{\mu\nu}-{P}^{\mu\lambda\rho\sigma}
{R}^\nu_{~\lambda\rho\sigma}
-2\nabla_\rho\nabla_\sigma{P}^{\rho\mu\nu\sigma}\big)
\mathcal{L}_\zeta{g}_{\mu\nu} \nn \\
&&+2\nabla_\mu\big[\zeta_\nu\big({P}^{\mu\lambda\rho\sigma}
{R}^{\nu}_{~\lambda\rho\sigma}-2\nabla_{\rho}\nabla_{\sigma}
{P}^{\rho\mu\nu\sigma}+{W}^{\mu\nu}\big)\big]
\, , \label{SumZiLieR}
\eea
which is reformulated in terms of both the Euler-Lagrange expression
${E}_{\mu\nu}$ and the surface term $\Theta^\mu$ with
$\delta\rightarrow\mathcal{L}_\zeta$ as
\bea
\sum^{m}_{i=0}
{Z}_{(i)}^{\lambda_1\cdot\cdot\cdot\lambda_i\mu\nu\rho\sigma}
\mathcal{L}_\zeta\nabla_{\lambda_1}
\cdot\cdot\cdot\nabla_{\lambda_i}{R}_{\mu\nu\rho\sigma}
&=&\left({E}_{\mu\nu}-\frac{\partial{L}}{\partial {g}^{\mu\nu}}
+\frac{1}{2}{L}{g}_{\mu\nu}\right)\mathcal{L}_\zeta{g}^{\mu\nu} \nn \\
&&+\nabla_\mu\Theta^\mu(\delta\rightarrow\mathcal{L}_\zeta)
\, . \label{SumZiLieR2}
\eea
Here we point out that Eq. (\ref{SumZiLieR2}) does not appear as a
consequence out of the imposition for the constraint that
$\Theta^\mu(\delta\rightarrow\mathcal{L}_\zeta)$ has to be
decomposed into the form (\ref{ThetaLie}) in advance. It is
attributed to the fact that adopting the definition for the Lie
derivative to straightforwardly compute the covariant divergence
of the surface term $\Theta^\mu$ given by Eq. (\ref{SurfT}) with
$\delta\rightarrow\mathcal{L}_\zeta$ results in the outcome
\be
\nabla_\mu\Theta^\mu(\delta\rightarrow\mathcal{L}_\zeta)
=2\nabla_\mu\big[\zeta_\nu\big({P}^{\mu\lambda\rho\sigma}
{R}^{\nu}_{~\lambda\rho\sigma}-2\nabla_{\rho}\nabla_{\sigma}
{P}^{\rho\mu\nu\sigma}+{W}^{\mu\nu}\big)\big]
=2\nabla_\mu\big(\zeta_\nu{X}^{\mu\nu}\big)
\,  \label{RelDivThetX}
\ee
that the quantity $\Theta^\mu(\delta\rightarrow\mathcal{L}_\zeta)$
is able to participate in Eq. (\ref{SumZiLieR2}).
Furthermore, note that the surface term $\Theta^\mu$ completely
originates from the sum of the scalar
${Z}_{(i)}^{\lambda_1\cdot\cdot\cdot\lambda_i\mu\nu\rho\sigma}
\delta\nabla_{\lambda_1}
\cdot\cdot\cdot\nabla_{\lambda_i}{R}_{\mu\nu\rho\sigma}$ with respect
to $i$ from $0$ to $m$ within Eq. (\ref{VariLagdens}). Based on this,
such a term, determined only by the variation of the Lagrangian
density, can be fully responsible for Eq. (\ref{RelDivThetX}).
Due to this, equation (\ref{RelDivThetX}) is valid identically
within the framework of the theories of gravity described by the
Lagrangian (\ref{LagCovR}), even within the case where the tensors
${Z}_{(i)}^{\lambda_1\cdot\cdot\cdot\lambda_i\mu\nu\rho\sigma}$s
are allowed to be arbitrary as shown by Eq. (\ref{ThetLieGFXQ0}).
This also indicates that we obtain
Eq. (\ref{RelDivThetX}) regardless of the existence or nonexistence
for either of Eqs. (\ref{LieLagdens}) and (\ref{DivJE}), although
equation (\ref{LieDLtoThet}) has provided a way to derive
Eq. (\ref{RelDivThetX}) in terms of the coexistence of the two
equations. What is more, equation (\ref{SumZiLieR2}) can be seen as
a consequence directly arising from the substitution of $\delta$
with $\mathcal{L}_\zeta$ in Eq. (\ref{SumZideltR}) below.

Actually, on the basis of the fact that the scalar $\sum^{m}_{i=0}
{Z}_{(i)}^{\lambda_1\cdot\cdot\cdot\lambda_i\mu\nu\rho\sigma}
\delta\nabla_{\lambda_1}\cdot\cdot\cdot\nabla_{\lambda_i}
{R}_{\mu\nu\rho\sigma}$ originates from the variation of the
Lagrangian density ${L}$, by utilizing the identity
(\ref{SumZiLieR}), we can offer another explanation for that
the coexistence of the two equations (\ref{LieLagdens}) and
(\ref{DivJE}) is a sufficient condition for giving rise to
Eq. (\ref{ThetaLie}) as shown by Eq. (\ref{LieDLtoThet}).
For the sake of doing so, direct computation shows that
the scalar $\sum^{m}_{i=0}
{Z}_{(i)}^{\lambda_1\cdot\cdot\cdot\lambda_i\mu\nu\rho\sigma}
\delta\nabla_{\lambda_1}\cdot\cdot\cdot\nabla_{\lambda_i}
{R}_{\mu\nu\rho\sigma}$
is associated with the covariant divergence of the surface
term $\Theta^\mu$ in the manner
\be
\sum^{m}_{i=0}
{Z}_{(i)}^{\lambda_1\cdot\cdot\cdot\lambda_i\mu\nu\rho\sigma}
\delta\nabla_{\lambda_1}
\cdot\cdot\cdot\nabla_{\lambda_i}{R}_{\mu\nu\rho\sigma}
=\left({E}_{\mu\nu}-\frac{\partial{L}}{\partial{g}^{\mu\nu}}
+\frac{1}{2}{L}{g}_{\mu\nu}\right)\delta{g}^{\mu\nu}
+\nabla_\mu\Theta^\mu
\, .  \label{SumZideltR}
\ee
Alternatively, equation (\ref{SumZideltR}) can be regarded as
the difference between Eqs. (\ref{VarLag}) and (\ref{VarLag3}).
Apart from this, under the replacement
$\delta\rightarrow\mathcal{L}_\zeta$, it is easy to check that
equation (\ref{SumZideltR}) accordingly turns into the difference
between
Eqs. (\ref{VarLag})$|_{\delta\rightarrow\mathcal{L}_\zeta}$
and (\ref{VarLag3})$|_{\delta\rightarrow\mathcal{L}_\zeta}$,
that is, the difference between Eqs. (\ref{LieVarLag}) and
(\ref{LieVarLag3}) below. Here the former of these two
equations is equivalent to Eq. (\ref{LieLagdens}) and the
latter is just Eq. (\ref{DivJE}). Furthermore, by means of
the comparison between the identity (\ref{SumZiLieR}) and
Eq. (\ref{SumZideltR})$|_{\delta\rightarrow\mathcal{L}_\zeta}$,
we obtain the covariant divergence of
$\Theta^\mu(\delta\rightarrow\mathcal{L}_\zeta)$ presented by
$\nabla_\mu\Theta^\mu(\delta\rightarrow\mathcal{L}_\zeta)
=2\nabla_\mu\big(\zeta_\nu{X}^{\mu\nu}\big)$, which is able to
give rise to Eq. (\ref{ThetaLie}). Hence, we arrive at the
conclusion that the combination rather than a single one of
Eqs. (\ref{LieLagdens}) and (\ref{DivJE}) can result in
Eq. (\ref{ThetaLie}) as well.

With Eq. (\ref{SumZiLieR2}) at hand, we concentrate on the relationship
between Eqs. (\ref{VarLag}) and (\ref{VarLag3}) after the variation
operator in both of them is replaced with the Lie derivative. Let us
start from Eq. (\ref{VarLag})$|_{\delta\rightarrow\mathcal{L}_\zeta}$,
that is,
\bea
\mathcal{L}_\zeta\mathcal{L}&=&\sqrt{-g}\left(\mathcal{L}_\zeta{L}
-\frac{1}{2}{L}{g}_{\mu\nu}\mathcal{L}_\zeta{g}^{\mu\nu}\right) \nn \\
&=&\sqrt{-g}\left[\Big(\frac{\partial{L}}{\partial{g}^{\mu\nu}}
-\frac{1}{2}{L}{g}_{\mu\nu}\Big)\mathcal{L}_\zeta{g}^{\mu\nu}
+\sum^{m}_{i=0}
{Z}_{(i)}^{\lambda_1\cdot\cdot\cdot\lambda_i\mu\nu\rho\sigma}
\mathcal{L}_\zeta\nabla_{\lambda_1}
\cdot\cdot\cdot\nabla_{\lambda_i}{R}_{\mu\nu\rho\sigma}\right]
\, . \quad \label{LieVarLag}
\eea
Substituting the identity (\ref{SumZiLieR2}) into the above equation
to cancel out the sum of the scalar
${Z}_{(i)}^{\lambda_1\cdot\cdot\cdot\lambda_i\mu\nu\rho\sigma}
\mathcal{L}_\zeta\nabla_{\lambda_1}
\cdot\cdot\cdot\nabla_{\lambda_i}{R}_{\mu\nu\rho\sigma}$
over $i$ leads to
\be
\mathcal{L}_\zeta\mathcal{L}=\sqrt{-g}\left[
{E}_{\mu\nu}\mathcal{L}_\zeta{g}^{\mu\nu}
+\nabla_\mu\Theta^\mu(\delta\rightarrow\mathcal{L}_\zeta)\right]
\, , \label{LieVarLag3}
\ee
which is just Eq. (\ref{VarLag3}) under the substitution
${\delta\rightarrow\mathcal{L}_\zeta}$. Conversely, equation
(\ref{LieVarLag}) can be derived out of Eq. (\ref{LieVarLag3}) by
utilizing the identity (\ref{SumZiLieR2}) once again to get rid
of the divergence term
$\nabla_\mu\Theta^\mu(\delta\rightarrow\mathcal{L}_\zeta)$
in Eq. (\ref{LieVarLag3}). Thus, we observe that the identity
(\ref{SumZiLieR2}) can be interpreted as a bridge establishing
the equivalence between Eqs. (\ref{LieVarLag}) and (\ref{LieVarLag3})
and equation (\ref{RelDivThetX}) can arise as the difference between
such two equations as illustrated in the above. The invalidity of
Eq. (\ref{RelDivThetX}) implies that this equivalence breaks down.
Moreover, we have the significant conclusion that
Eq. (\ref{RelDivThetX}) is definitely valid as long as the variation
operator $\delta$ in Eqs. (\ref{VarLag}) and (\ref{VarLag3}) is
allowed to be replaced with the Lie derivative $\mathcal{L}_\zeta$.

As a natural consequence of the relation between Eqs. (\ref{LieVarLag})
and  (\ref{LieVarLag3}), we arrive at the conclusion that the equivalence
relationship between Eqs. (\ref{VarLag}) and (\ref{VarLag3}), denoted by
\be
\text{Equation }(\ref{VarLag})
\quad\Leftrightarrow\quad
\text{Equation }(\ref{VarLag3})
\, , \label{EquiRelVaL}
\ee
causes their equivalence with the variation operator $\delta$ substituted
by the Lie derivative $\mathcal{L}_\zeta$, that is,
\be
\text{Equation }(\ref{VarLag})|_{\delta\rightarrow\mathcal{L}_\zeta}
\quad\Leftrightarrow\quad
\text{Equation }(\ref{VarLag3})|_{\delta\rightarrow\mathcal{L}_\zeta}
\, . \label{LieLEJequiR3}
\ee
Furthermore, attributed to the fact that the two equations
(\ref{LieLagdens}) and (\ref{DivJE}) are completely equivalent to
Eqs. (\ref{VarLag}) and (\ref{VarLag3})
under $\delta\rightarrow\mathcal{L}_\zeta$, respectively, both
of them naturally inherit the equivalence relationship displayed by
Eq. (\ref{LieLEJequiR3}). Due to this, the existence for either of
Eqs. (\ref{LieLagdens}) and (\ref{DivJE}) implies their coexistence,
while equation (\ref{ThetaLie}) or (\ref{RelDivThetX}) is a necessary
condition for the coexistence of both the two equations.

At the end of this appendix, the relations among the three equations
(\ref{RelDivThetX}), (\ref{LieVarLag}) and (\ref{LieVarLag3}) are
summarized as
\bea
\text{Equations }(\ref{RelDivThetX})\text{ \& }
(\ref{LieVarLag})\quad&\Rightarrow&\quad
\text{Equation }(\ref{LieVarLag3}) \, , \nn \\
\text{Equations }(\ref{RelDivThetX})\text{ \& }
(\ref{LieVarLag3})\quad&\Rightarrow&\quad
\text{Equation }(\ref{LieVarLag}) \, , \nn \\
\text{Equations }(\ref{LieVarLag})\text{ \& }
(\ref{LieVarLag3})\quad&\Rightarrow&\quad
\text{Equation }(\ref{RelDivThetX})
\, . \label{RelDivThLieVLag}
\eea
By utilizing the equivalence relationships
\bea
\text{Equation }(\ref{ThetaLie})
\quad&\Leftrightarrow&\quad
\text{Equation }(\ref{RelDivThetX}) \, , \nn \\
\text{Equation }(\ref{LieLagdens})
\quad&\Leftrightarrow&\quad
\text{Equation }(\ref{LieVarLag}) \, , \nn \\
\text{Equation }(\ref{DivJE})
\quad&\Leftrightarrow&\quad
\text{Equation }(\ref{LieVarLag3})
\, , \label{EquiRTieLV}
\eea
equation (\ref{RelDivThLieVLag}) enables us to arrive at the conclusion
that the arbitrary two equations among
the three ones (\ref{ThetaLie}), (\ref{LieLagdens}) and (\ref{DivJE})
can jointly serve as the sufficient condition for producing the
remaining one. Such a conclusion has been presented in terms of the
combination of Eqs. (\ref{LieLEJequiR}) and (\ref{LieDLtoThet}) within
Section \ref{five}.

\end{document}